\documentclass{aa}  

\usepackage{graphicx}
\usepackage{txfonts}

\usepackage{bm}
\usepackage{upgreek}
\usepackage{fancyvrb}

\newcommand{\lin}[0]{_{\mathrm{in}}}
\newcommand{\lout}[0]{_{\mathrm{out}}}
\newcommand{\dd}{\,\mathrm{d}}
\newcommand{\brac}[1]{\left\langle #1 \right\rangle}

\newcommand{\Rm}[0]{\mathcal{R}_m}

\graphicspath{ {Pics/} }

\begin{document} 
\VerbatimFootnotes

\title{Gravitoturbulent dynamo in global simulations of gaseous disks}

\author{William B\'ethune\inst{1} \and Henrik Latter\inst{2}}
\authorrunning{W. B\'ethune \& H. Latter}

\institute{Institut f\"ur Astronomie und Astrophysik, Universit\"at T\"ubingen, Auf der Morgenstelle 10, 72076 T\"ubingen, Germany
  \and
  DAMTP, University of Cambridge, CMS, Wilberforce Road, Cambridge, CB3 0WA, UK\\
  \email{william.bethune@uni-tuebingen.de, hl278@cam.ac.uk}
}

\date{Received ****; accepted ****}

\abstract
    {The turbulence driven by gravitational instabilities (GIs) can amplify magnetic fields in massive gaseous disks. This GI dynamo may appear in young circumstellar disks, whose weak ionization challenges other amplification routes, as well as in active galactic nuclei. Although regarded as a large-scale dynamo, only local simulations have so far described its kinematic regime.}
    {We study the GI dynamo in global magnetohydrodynamic (MHD) models of accretion disks, focusing on its kinematic phase.}
    {We perform resistive MHD simulations with the \textsc{Pluto} code for different radiative cooling times and electrical resistivities. A weak magnetic field seeds the dynamo, and we adopt mean-field and heuristic models to capture its essence.}
    {We recover the same induction process leading to magnetic field amplification as previously identified in local simulations. The dynamo is, however, global in nature, connecting distant annuli of the disk via a large-scale dynamo mode of a fixed growth rate. This large-scale amplification can be described by a mean-field model that does not rely on conventional $\alpha$-$\Omega$ effects. When varying the disk parameters we find an optimal resistivity that facilitates magnetic amplification, whose magnetic Reynolds number, $\Rm \lesssim 10$, is substantially smaller than in local simulations. Unlike local simulations, we find an optimal cooling rate and the existence of global oscillating dynamo modes. The nonlinear saturation of the dynamo puts the disk in a strongly magnetized turbulent state on the margins of the effective range of GI. In our simulations, the accretion power eventually exceeds the threshold required by local thermal balance against cooling, leaving the long-term nonlinear outcome of the GI dynamo uncertain.}
    {}

\keywords{Accretion, accretion disks -- Dynamo -- Gravitation -- Magnetohydrodynamics (MHD) -- Turbulence}

\maketitle

\section{Introduction}

Magnetic fields play a central role in the evolution of accretion disks. They are often considered as essential for mass accretion due to their ability to power jets and disk winds \citep{blandford82,pelletier92} and to trigger turbulence via the magnetorotational instability \citep[MRI;][]{balbus91}. Dynamically consequent magnetic fields may be supplied by infalling plasma during the formation of the disk \citep[protostellar ones, for example; see][]{zhao20} or by the overflow of a binary companion \citep{ju16,ju17,pjanka20}, and their accumulation can even lead to magnetically dominated flows \citep[e.g.,][]{tchekho11}. Alternatively, magnetic fields may be amplified and sustained by a dynamo process operating inside the disk. 

Dynamos associated with the MRI have received the most attention due to their ubiquity in astrophysical disks and their relative strength \citep{hawley92,hawley96,rincon07,lesur08,gressel15,walker16,riols17a,mama20}. Alternative fluid dynamos exist\footnote{For an example of wave-induced dynamo, see \citet{vishniac90}.} and become especially important when the ionization fraction is too low to sustain the MRI \citep[e.g.,][]{kawasaki21}. Recently, turbulence caused by the gravitational instability \citep[GI;][]{toomre64} of massive disks has offered a novel and compelling way to amplify magnetic fields via the so-called GI dynamo \citep{riols19}, not least because it can outcompete the MRI in certain regimes \citep{riols18a}.

The conditions for GI are expected to be met in young circumstellar disks \citep{durisen07,kratter16} and active galactic nuclei \citep[AGN;][]{shlosman87,goodman03}. The instability leads to a self-sustained turbulent state if radiative cooling is sufficiently slow \citep{gammie01,rice03,lin16,booth19}, at which point it facilitates angular momentum transport via spiral density and gravitational perturbations \citep{paczynski78,linpringle87,balbuspap99}. Because the GI develops in the plane of the disk, much hydrodynamical work has excluded the out-of-plane dynamics outright by employing two-dimensional models, or otherwise neglected its analysis \citep[exceptions include][]{boley06a,shi14,riols17,bethune21}. However, understanding the three-dimensional (3D) structure of the flow is absolutely crucial when assessing the dynamo question. 

A number of global magnetohydrodynamic (MHD) simulations have described the susceptibility of magnetized disks to fragmentation \citep{fromang05,forgan17,zhao18,wurster19} or the interplay of the MRI with the GI in disks with a preexisting strong magnetization \citep{fromang04a,fromang04b,deng20}. But to date, only the local (shearing box) simulations of \citet{riols18a,riols19} have probed the dynamo properties of GI turbulence at low magnetization and in regimes excluding the MRI, thus witnessing the kinematic phase of the GI dynamo and thereby determining its underlying mechanism. However, a problem they faced was the emergence of large-scale magnetic fields --- relative to the largest turbulent eddies, and indeed the computational domain --- which called into question the local approximation itself. It is likely that global (geometry-dependent) effects are important and need to be accounted for.

In this paper, we combine the global and the kinematic through MHD simulations of gravitoturbulent disks, thereby presenting clean results on the global GI dynamo process. Unlike previous global studies, we start with magnetic fields that are sufficiently weak such that the MRI is initially ineffective, especially when Ohmic diffusion is added. We use weighted averages to separate the GI dynamo into radially local and global parts. The local part helps us make contact with shearing box results and mean-field dynamo models, whereas the global part is entirely novel and comprises one of the main achievements of the paper. Our focus is on the kinematic dynamo regime, but we present some results on its saturated phase that reinforce the idea that the saturation route ends in a highly magnetized state (plasma $\beta$ of order unity). 

The rest of the paper is subdivided into seven sections. We present our framework in Sect. \ref{sec:method}, including the chosen disk model, numerical scheme, and data reduction techniques. Section \ref{sec:basics} provides an overview of the disks in which the dynamo operates, including descriptions of the mean flow and magnetic field. In Sect. \ref{sec:kinematic} we decompose the kinematic induction loop into its components and then reconstruct a local geometric interpretation of the dynamo mechanism. We then show how global behaviors stem from the radial diffusion of magnetic energy. In Sect. \ref{sec:mfd} we establish a reduced dynamo model based on the local mean fields and whose parameters are directly measured from simulations. Section \ref{sec:params} undertakes a survey of various disk parameters to assess the robustness of our previous results. We examine in Sect. \ref{sec:saturated} the saturation of the GI dynamo as observed in our simulations. Finally, we summarize our results and some of their possible extensions in Sect. \ref{sec:theend}.

\section{Method} \label{sec:method}

Our numerical setup is nearly identical to that of \citet{bethune21}, the main novelty being the inclusion of a magnetic field. Below we recall the properties of the adopted disk model and of its implementation within the grid-based and shock-capturing code \textsc{Pluto}. We also define conventions and notations for various quantities used throughout the paper.

\subsection{Global disk model and units}

We simulated a disk of self-gravitating fluid orbiting a more massive central object. We used both spherical coordinates $\left(r,\theta,\varphi\right)$ and cylindrical coordinates $\left(R,\varphi,z\right)$ with the central object at the origin and the rotation axis of the disk as the polar (resp. vertical) direction. We considered only a finite radial extent $\left[r\lin,r\lout\right]$, excluding the central region. For simplicity, we assumed that this frame is inertial by neglecting the reaction of the central object to the gravity of the disk, which is a reasonable approximation for the relatively low disk masses considered \citep{heemskerk92,michael10}. We modeled radiative energy losses via a simple cooling function that brings the gas temperature toward zero (in the absence of heating) over a prescribed number of orbits. Finally, we assumed that the disk is electrically conducting such that fluid motions can induce and react to magnetic fields according to a resistive MHD description. 

Throughout this paper, we take the mass $m_{\star}$ of the central object as the constant mass unit and denote $M \equiv m_{\mathrm{disk}}/m_{\star}$ the initial mass of the disk. We use the inner radius $r\lin$ of the simulation domain as the constant length unit. After absorbing the gravitational constant $G$ into the masses, $\Omega_{\star} \equiv \sqrt{m_{\star}/R^3}$ denotes the Keplerian frequency at radius $R$, and we take $\Omega\lin \equiv \Omega_{\star}\left(r\lin\right)$ as the inverse time unit. The inner Keplerian period is denoted $t\lin \equiv 2\uppi/\Omega\lin$. Magnetic fields are measured by their corresponding Alfv\'en velocities when the gas density $\rho=1$ in these units.

\subsection{Governing equations} \label{sec:goveq}

The equations governing the evolution of the gas density $\rho$, velocity $\bm{V}$, pressure $P$, and magnetic field $\bm{B}$ are:
\begin{align}
  \partial_t \rho &= -\nabla \cdot \left( \rho \bm{V}\right), \label{eqn:dtrho}\\
  \rho \left(\partial_t \bm{V} + \bm{V}\cdot \nabla \bm{V}\right) &= -\nabla P - \rho \nabla \Phi + \bm{J}\times \bm{B}, \label{eqn:dtv}\\
  \partial_t P + \bm{V}\cdot\nabla P &= -\gamma P\, \nabla\cdot\bm{V} + \left(\gamma-1\right) \eta J^2 + \Lambda, \label{eqn:dtP}\\
  \partial_t \bm{B} &= \nabla \times \left( \bm{V} \times \bm{B} - \eta \bm{J}\right). \label{eqn:dtB}  
\end{align}
We used a constant adiabatic exponent $\gamma=5/3$ to facilitate comparison with \citet{riols19}, \citet{deng20}, and \citet{bethune21}. The gravitational potential $\Phi$ is the sum of a constant central potential $-m_\star/r$ and a time-dependent contribution from the disk satisfying Poisson's equation $\Delta\Phi_\mathrm{disk} = 4\uppi \rho$. The cooling function in Eq. \eqref{eqn:dtP} takes the form $\Lambda\simeq -\left(\Omega_{\star}/\tau\right) P$, parametrized by a constant dimensionless cooling time\footnote{We denote the dimensionless cooling time by $\tau$, as \citet{gammie01} and \citet{riols19}, in place of $\beta$ used by \citet{bethune21}.} $\tau$. The magnetic field influences the gas's momentum via the Lorentz force in Eq. \eqref{eqn:dtv} and its internal energy via Joule heating in Eq. \eqref{eqn:dtP}, where $\bm{J} \equiv \nabla\times{\bm{B}}$ is the electric current density. For later reference, we denote by $\bm{\mathcal{E}}\equiv -\bm{V}\times\bm{B}$ the ideal electromotive force (EMF) in Eq. \eqref{eqn:dtB}. We omitted explicit viscosity in Eqs. \eqref{eqn:dtv} \& \eqref{eqn:dtP} and relied mainly on shocks to thermalize kinetic energy.

We modeled the finite electric conductivity of the gas with an Ohmic resistivity $\eta$ in Eqs. \eqref{eqn:dtP} \& \eqref{eqn:dtB}, which we parametrized by a magnetic Reynolds number $\Rm = \Omega h^2 / \eta$. Since the characteristic thickness $h$ of the disk can vary with its thermal stratification and self-gravity, there are multiple possible definitions for $\Rm$. For simplicity, we used the initial profile of the pressure scale height, $h_\mathrm{init}/R\equiv\left(c_{s}/V_{\varphi}\right)_\mathrm{init}$, given the isothermal sound speed $c_{s} \equiv \sqrt{P/\rho}$. The resistivity is therefore constant in time according to the prescribed value of $\Rm \equiv \Omega_{\star} h_{\mathrm{init}}^2 / \eta$. The actual ratio $\Omega h^2 / \eta$ may vary with radius and time, but we checked that the characteristic scale height $h_\rho$ (see Sect. \ref{sec:datared} below) remains within ten per cent of the initial hydrostatic scale height $h_{\mathrm{init}}$ on average, and they differ by at most 30 per cent locally.

\subsection{Numerical integration scheme}

We used the finite volume code \textsc{Pluto} version 4.3 \citep{mignone07} to integrate Eqs. \eqref{eqn:dtrho}-\eqref{eqn:dtB} in conservative form on a static spherical grid. The computational domain $\left(r/r\lin,\theta,\varphi\right) \in \left[1,32\right]\times \left[\uppi/2\pm 0.35\right] \times \left[0,2\uppi\right]$ was meshed with $518 \times 96 \times 512$ grid cells, with a logarithmic sampling in radius and a finer meridional resolution near the disk midplane ($\theta \in \uppi/2 \pm 0.1$).

As \citet{bethune21}, we used a second-order Runge-Kutta time stepping with CFL coefficient $0.3$, and a piecewise linear reconstruction \citep{mignone14} of the primitive variables at cell interfaces with the symmetric slope limiter of \citet{vanleer74}. Interface fluxes were computed with the HLLd approximate Riemann solver of \citet{miyoshi05}. At strong discontinuities, we switched to the more dissipative MINMOD slope limiter \citep{roe86} and to a two-state HLL Riemann solver \citep{harten83}. The magnetic field obeys the constrained transport formalism \citep{evans88} so that $\nabla\cdot\bm{B}=0$ is guaranteed from the start down to machine precision. We used the \verb|UCT_HLL| method to reconstruct the EMF on cell edges \citep{zanna03,londrillo04}. Unacceptable magnetic energy injection occurred at the domain boundaries when using other reconstruction schemes (see Appendix \ref{app:uct}). Ohmic resistivity was integrated explicitly via super-time-stepping \citep{alexiades96}, although only the most resistive case ($\Rm=1$) benefited from the associated acceleration. 

\subsection{Initial and boundary conditions}

The initial conditions for the hydrodynamic variables are the same as those of \citet{bethune21}. The gas initially rotates at $V_{\varphi} = \sqrt{\left[m_{\star}+m_{\mathrm{disk}}\left(R\right)\right]/R}$, accounting for the enclosed disk mass $m_{\mathrm{disk}}\left(R\right) = \int_{r\lin}^{R} 2\uppi \Sigma x \dd x$. The initial surface density $\Sigma \propto R^{-2}$ was scaled by the chosen disk mass. The initial sound speed $c_s \propto R^{-1/2}$ gives a roughly constant aspect ratio $h/R$, and it was adjusted so that the Toomre number $Q \simeq \Omega_{\star} c_{s} / \uppi \Sigma \approx 1$.

We initialized the magnetic field with a net toroidal component $B_{\varphi}^2 = 10^{-12} P$ confined inside $1.1r\lin < r < 0.9 r\lout$ and $\left\vert \theta - \uppi/2\right\vert < 0.1$. This magnetic field is present from the start of our simulations, before the onset of GI turbulence. Compared to the expected dynamo growth times, the initial transition to a quasi-steady turbulent state is short and may be neglected. Given how weak the initial magnetic field is, the MRI would operate appreciably for wavenumbers $k h_\mathrm{init} \sim 10^6$ in ideal MHD \citep{balbus92,ogilvie96}. Such scales are much smaller than our finest grid increment, and numerical diffusion would easily overcome the MRI at our grid scale. The addition of a finite Ohmic resistivity guarantees that the MRI cannot operate before reaching an average $B^2/P \gtrsim 10^{-2}$.

The boundary conditions for the hydrodynamic variables $\rho$, $\bm{V}$, and $P$ are the same as those of \citet{bethune21}. We set the tangential components of the magnetic field to zero in the ghost cells of both the radial ($r$) and meridional ($\theta$) boundaries. When a grid cell adjacent to a boundary supports a nonzero magnetic flux, the component normal to the boundary is automatically determined by $\nabla\cdot\bm{B}=0$ in the ghost cells. 

\subsection{Data reduction} \label{sec:datared}

Throughout this paper, the word ``mean'' designates the result of some spatial integration, and fluctuations are defined relative to this average. It is conceptually different from separating the flow into a turbulent component over a quiescent background, as is customary in classic mean-field theory. This distinction should be kept in mind later when we discuss mean flows that originate in turbulent motions, although the average of most quantities does reflect their location and time-independent properties.

Our simulation diagnostics rely on several averaging procedures to extract clear signals out of turbulence. We use angled brackets $\brac{\bullet}_{x_i}$ to denote spatial averages over the dimensions $x_i$, omitting the subscripts when unnecessary. We use the subscript $\rho$ to denote a density-weighted meridional and azimuthal average. Since we focus on thin disks, the meridional ($\theta$) and vertical ($z$) denominations may be used interchangeably. 

We used radial averages to maximize the statistical quality of our diagnostics and to extract vertical structures common at all radii. They covered the interval $r/r\lin \in \left[2,8\right]$ and were evaluated with a $\dd \log r$ measure corresponding to our grid spacing. To cancel the global radial gradients in the disk, we rescaled some quantities prior to their averaging. We used the density-weighted angular velocity $\Omega_\rho \equiv \brac{V_{\varphi}/R}_{\rho}$ and isothermal sound speed $c_\rho \equiv \sqrt{\brac{P}_{\theta,\varphi} / \brac{\rho}_{\theta,\varphi}}$ to define radial profiles used as averaging weights: a typical thickness $h_\rho \equiv c_\rho / \Omega_\rho$, a typical velocity $\bar{V} \equiv \Omega_\rho R$, and a typical magnetic field strength $\bar{B} \equiv \sqrt{\brac{B^2}_{\theta,\varphi}}$. We then obtained the mean meridional profiles of the reduced velocity $\bm{v}$, mass flux $\bm{q}$, magnetic field $\bm{b}$, electric current density $\bm{j}$, and ideal EMF $\bm{\epsilon}$ as:
\begin{equation} \label{eqn:redvar}
  \left(\bm{v}, \bm{q}, \bm{b}, \bm{j}, \bm{\epsilon} \right) \equiv \brac{\left(  \frac{\bm{V}}{\bar{V}} , \frac{\rho\bm{V}}{\Sigma \Omega_\rho} ,   \frac{\bm{B}}{\bar{B}} , \frac{\bm{J}}{\bar{B}/h_\rho} , \frac{\bm{\mathcal{E}}}{\bar{V}\bar{B}}  \right)}_{\varphi, r}.
\end{equation}
Each dimensionless ratio on the right-hand side effectively fluctuates about the average on the left-hand side by a factor of a few at all radii and azimuths. To produce converged meridional profiles, we further averaged them over time. The results are smooth functions of latitude ($\theta$) representing global and persistent background structures embedded in turbulent fluctuations.

\section{Overview of the kinematic dynamo phase} \label{sec:basics}

\begin{figure}
  \centering
  \includegraphics[width=\columnwidth]{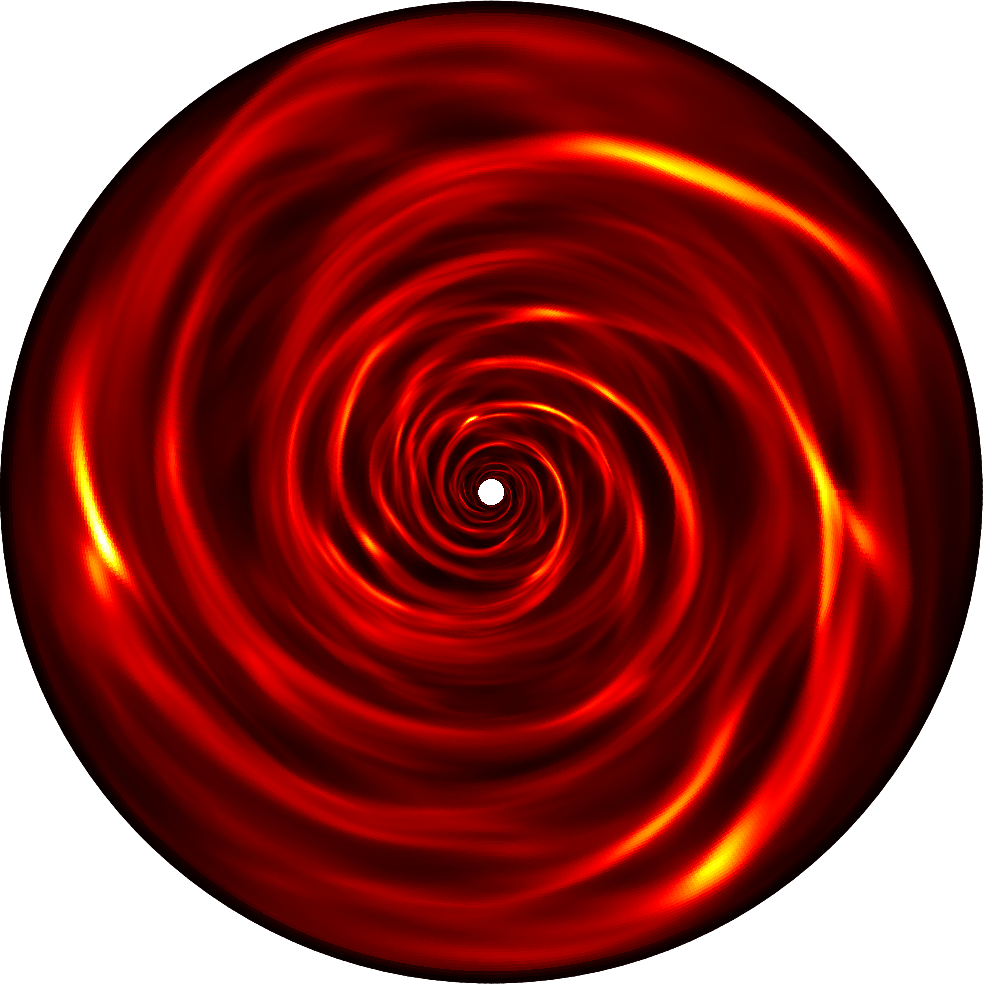}
  \caption{Flattened distribution of the gas surface density $R^2\Sigma$ at time $400t\lin$ over the whole domain of run M3T10R10; colors range linearly from zero to three times the median of $R^2\Sigma$.}
  \label{fig:m3b10r10t400_EQt_sigR2}
\end{figure}

In this section, we describe the broad properties of the turbulent disk and dynamo in a simulation chosen as reference; other disk parameters will be considered in Sect. \ref{sec:params}. With a disk mass $M=1/3$, a dimensionless cooling time $\tau=10$, and a magnetic Reynolds number $\Rm = 10$, we label this run M3T10R10. It has the same disk mass and cooling time as the reference simulation presented by \citet{bethune21}, and its moderate magnetic Reynolds number places it in the kinematic dynamo regime identified by \citet[][see their figure 2]{riols19}. 

\subsection{Global, secular growth of magnetic energy} \label{sec:magnetup}

\begin{figure}
  \centering
  \includegraphics[width=\columnwidth]{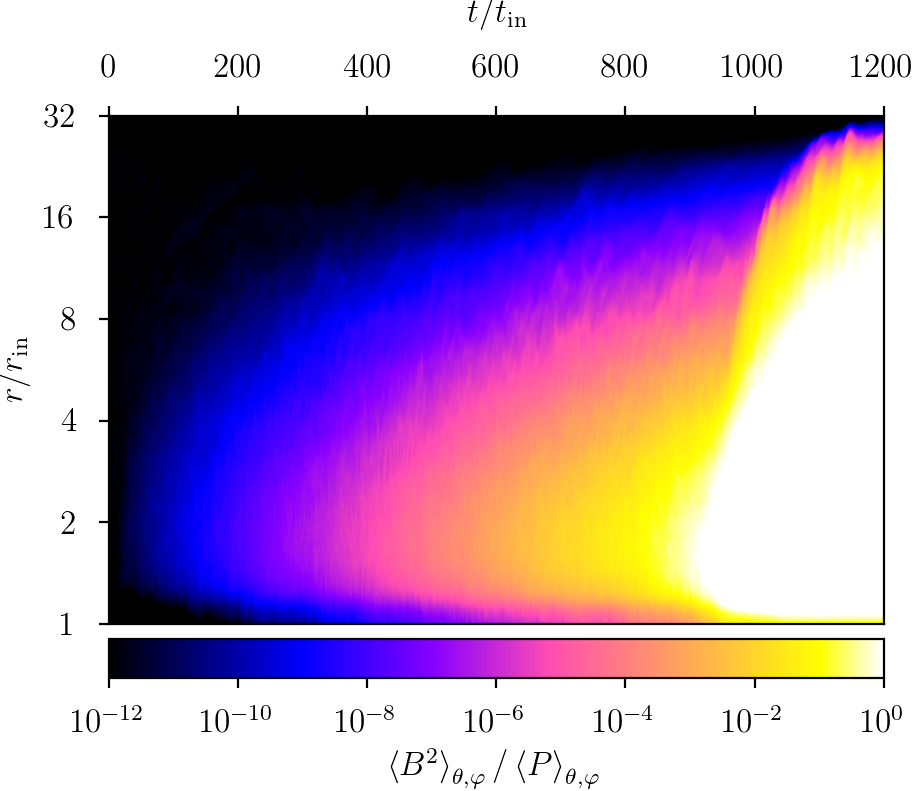}
  \caption{Evolution in time (horizontal axis) of the radial profile (vertical axis) of the disk magnetization $B^2/P$ in run M3T10R10.}
  \label{fig:m3b10r10_map_b2}
\end{figure}

Turbulence takes about $200 t\lin$ to settle over the whole domain. From this point on, velocity fluctuations evolve on local orbital times and on lengthscales comparable to the local disk thickness, as illustrated in Fig. \ref{fig:m3b10r10t400_EQt_sigR2}. Meanwhile, Fig. \ref{fig:m3b10r10_map_b2} shows the dimensionless magnetization ratio $B^2/P$ exhibiting a large-scale envelope that grows by 12 orders of magnitude over $900t\lin$. 

Noting that $P=\rho c_s^2$ and that the disk supports sonic turbulence, the Alfv\'en velocity $V_{\mathrm{A}}\equiv \sqrt{B^2/\rho}$ remains negligible compared to both the velocity fluctuations and the sound speed as long as $B^2/P \ll 1$. In this regime, both the Lorentz force and Joule heating are unimportant, and hence the magnetic induction Eq. \eqref{eqn:dtB} reduces to a linear initial-value problem for $\bm{B}$ in a given velocity field $\bm{V}$. We focus here on this kinematic dynamo phase and postpone the dynamo saturation ($B^2/P \gtrsim 1$) to Sect. \ref{sec:saturated}.

\begin{figure}[t]
  \centering
  \includegraphics[width=\columnwidth]{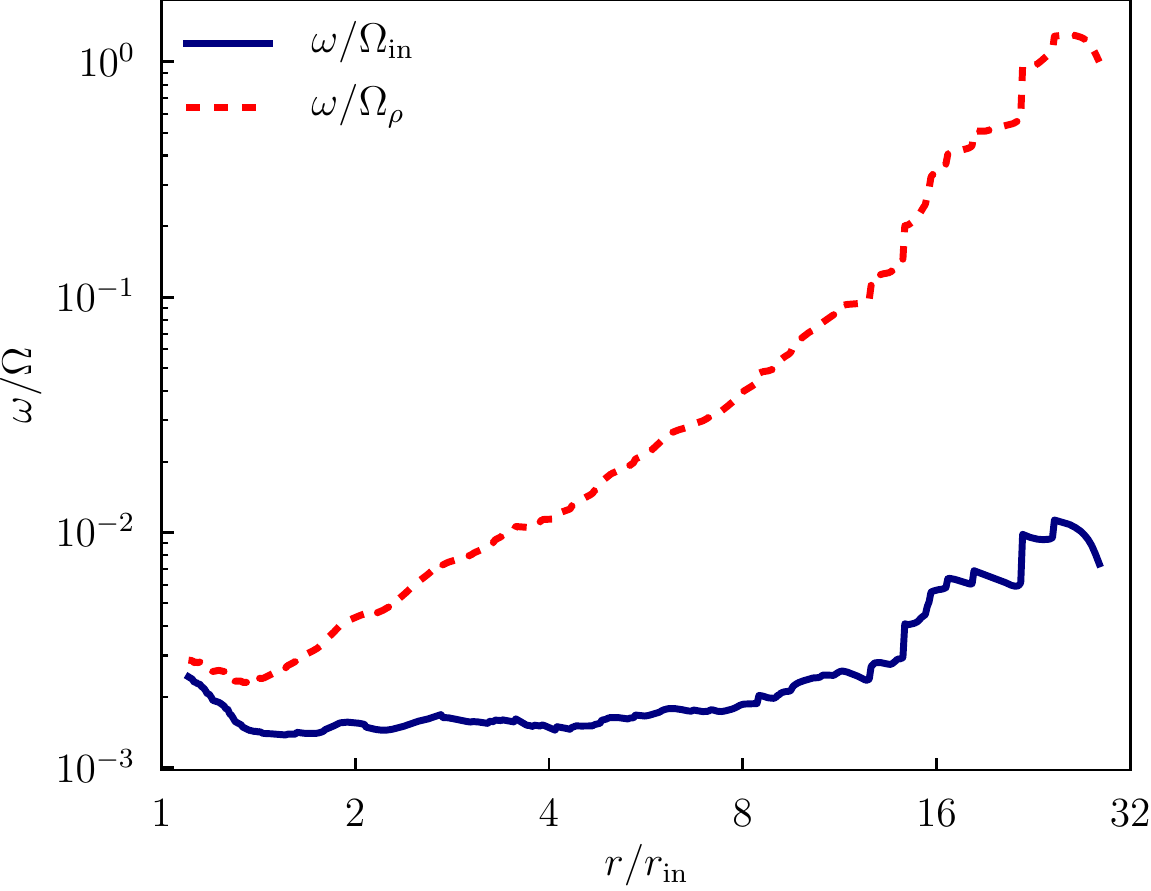}
  \caption{Growth rate of the magnetic field amplitude $\bar{B}\equiv\sqrt{\brac{B^2}_{\theta,\varphi}}$ as a function of radius in run M3T10R10, normalized either by the (fixed) inner Keplerian frequency $\Omega\lin$ (\emph{solid blue}) or by the (radially varying) density-weighted orbital frequency $\Omega_{\rho}$ (\emph{dashed red}). Only the inner parts $r/r\lin \lesssim 12$ have reached a steady exponential growth phase.}
  \label{fig:m3b10r10_rad_growth}
\end{figure}

As long as $\bm{B}$ is dynamically unimportant, its amplitude $\bar{B}$ should vary exponentially in time. We measured the evolution of $\bar{B}$ during one outer orbit $\approx 181t\lin$ before reaching $B^2/P=10^{-2}$ at any given radius and observed exponential growth. We plot in Fig. \ref{fig:m3b10r10_rad_growth} the radial profile of the growth rate $\omega$ estimated at each radius by fitting the slope of $\frac{1}{2}\log\left(\bar{B}^2\right)$ as a function of time. 

The solid blue curve in Fig. \ref{fig:m3b10r10_rad_growth} indicates that the growth rate $\omega \approx 1.6 \times 10^{-3} \Omega_{\lin}$ is roughly constant up to $r/r\lin \lesssim 12$. Beyond $r/r\lin \gtrsim 12$ our measurements of $\omega$ become inaccurate because the chosen time interval covers only a few orbital periods, and the dynamo has not settled into a steady growth yet at these radii. If instead we measure the growth rate relative to the local orbital frequency $\Omega_{\rho}$, we obtain a steady increase with radius up to $\omega / \Omega_\rho \sim 1$, as shown by the dashed red curve.

Figure \ref{fig:m3b10r10_rad_growth} reveals that magnetic amplification proceeds with a single growth rate irrespective of the local dynamical time. While significant growth takes hundreds of orbits in the inner parts, it happens on nearly orbital times in the outer parts. Thus, the dynamo mode causally connects distant radii, and its growth is likely limited by its outer extent (see Sect. \ref{sec:global}). Figure \ref{fig:m3b10r10_rad_growth} demonstrates a global property of the GI dynamo that shearing box simulations cannot capture. Because the dynamo mode in Fig. \ref{fig:m3b10r10_map_b2} exhibits lengthscales much longer than the turbulent driving scale, we also conclude that it is a large-scale mean-field dynamo --- whose type we determine in Sects. \ref{sec:kinematic} \& \ref{sec:mfd}. 

\subsection{Hydrodynamic (GI) turbulence} \label{sec:hydroturb}

\begin{figure}
  \centering
  \includegraphics[width=\columnwidth]{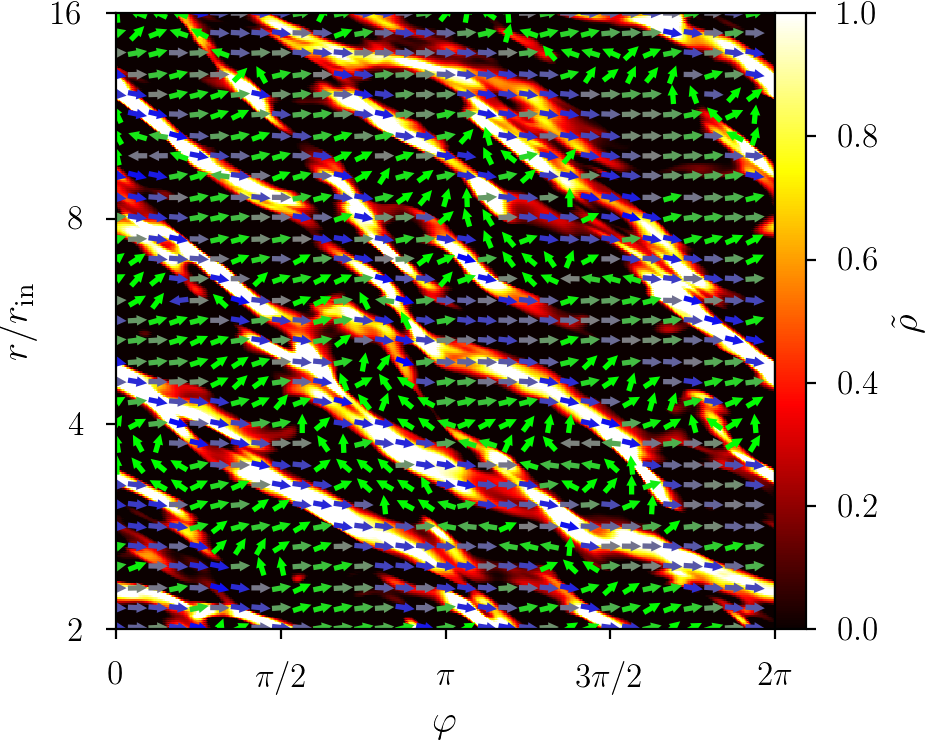}
  \caption{Equatorial slice in run M3T10R10 at time $400 t\lin$, zoomed in on the radial interval $r/r\lin \in \left[2,16\right]$. The color map shows the relative density fluctuations as defined by Eq. \eqref{eqn:relrho} while the arrows show the orientation of the magnetic field sampled in the midplane, colored in blue when inward ($B_r<0$) or green when outward ($B_r>0$).}
  \label{fig:m3b10r10t400_EQt_ddn_B}
\end{figure}

\begin{figure*}
  \centering
  \includegraphics[width=\textwidth]{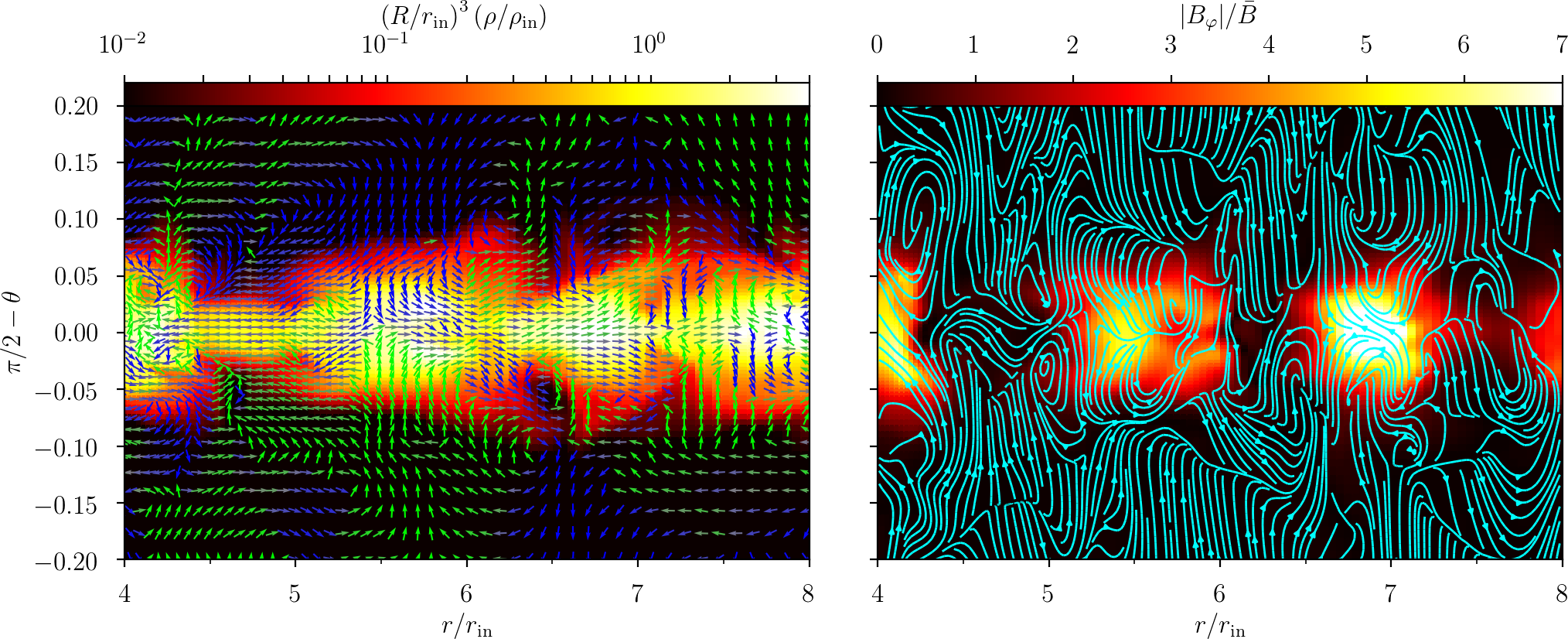}
  \caption{Meridional slice at time $400 t\lin$ and angle $\varphi=3\uppi/8$ in run M3T10R10. \emph{Left panel:} Density distribution flattened by a factor of $\left(R/r\lin\right)^3$ (color scale) and orientation of the poloidal velocity field, green (resp., blue) being oriented upward (resp., downward). \emph{Right panel:} Toroidal magnetic field relative to the local mean (color scale) and integral curves of the poloidal magnetic field. The density is maximal at $r/r\lin\approx 4.0, 5.6$, and $6.9$.}
  \label{fig:m3b10r10t400_RZCut_phi192_r4_VB}
\end{figure*}

In this subsection, we describe the small-scale features of the hydrodynamical turbulence that are important for the morphology and growth of the dynamo field. Figure \ref{fig:m3b10r10t400_EQt_ddn_B} shows the relative surface density fluctuations
\begin{equation} \label{eqn:relrho}
  \tilde{\rho} \equiv \frac{ \brac{\rho - \brac{\rho}_{\theta,\varphi}}_{\theta}}{\brac{\rho}_{\theta,\varphi}}
\end{equation}
at time $400 t\lin$ in the equatorial plane of run M3T10R10. This is similar to Fig. \ref{fig:m3b10r10t400_EQt_sigR2} after unfolding the azimuthal direction. The bright stripes where the density is maximal correspond to spiral wakes. Because of our choice of initial conditions, these wakes share a similar pitch angle ($\mathrm{d}\log R / \mathrm{d} \varphi$) at all radii, thus generating logarithmic spiral arms globally. Due to the relatively short cooling time $\tau=10$, the GI sustains vigorous turbulence and sharp density contrasts \citep[e.g.,][]{cossins09}.

To reveal the 3D nature of the flow, we sampled meridional slices of the disk at different angles $\varphi$. The left panel of Fig. \ref{fig:m3b10r10t400_RZCut_phi192_r4_VB} shows the density and poloidal velocity components in the $\varphi=3\uppi/8$ plane and $r/r\lin \in \left[4,8\right]$ interval. The velocity field roughly displays a mirror symmetry about the midplane: the gas simultaneously rises or falls on both sides at any given radius. This symmetry hints at a connection between the midplane disk dynamics and the vertical circulations developing in its corona. The density maxima located at $r/r\lin\approx 4.0$, $5.6$, and $6.9$ correspond to spiral wakes that are surrounded by varied flow patterns. At $r/r\lin \approx 4$ the velocity converges toward the wake in the midplane and splashes vertically away from it. At $r/r\lin \approx 5.6$ we see the opposite trend: the gas falls vertically toward the wake and evacuates it in the midplane. At $r/r\lin \approx 6.9$ the gas is lifted vertically as it moves radially past the wake. 

Distinct poloidal rolls appear at $\left(r/r\lin,\,\uppi/2-\theta\right) \approx \left(4.5,0.05\right)$, $\left(6.4,-0.1\right)$, and $\left(6.9,0.05\right)$, yet we had difficulty identifying a consistent quadrupolar pattern encompassing wakes, as witnessed in the local simulations of \cite{riols19} and \cite{riols21}. Instead, such circulations correlate with the gas density in only an average sense, as shown by \citet[][figure 14]{bethune21}. The discrepancy with local simulations may issue from our different vertical stratifications (see Sect. \ref{sec:thermo}) or from the symmetry constraints (periodicity and zero net momentum) enforced in shearing boxes. The variety of poloidal flow structures may also reflect different evolutionary stages of the wakes. \citet[][figure 12]{bethune21} showed that these wakes are transient and survive only for a fraction of local orbital time. One possible way to envision their destruction is to simply reverse the orientation of the surrounding flow.

\subsection{Small-scale magnetic field structure}

The velocity fluctuations responsible for producing spiral density maxima also distort the magnetic field on scales much smaller than the global envelope described in Sect. \ref{sec:magnetup}. The arrows in Fig. \ref{fig:m3b10r10t400_EQt_ddn_B} represent the orientation of the magnetic field sampled in the equatorial plane. Most arrows point to the right (increasing $\varphi$), meaning that the toroidal component $B_{\varphi} > 0$ has a prefered sign in the disk midplane. The radial component $B_r$ is typically negative inside spiral wakes and positive in the low-density region between wakes, in agreement with local simulations \citep[][]{riols19}. While in-plane motions cannot lead to dynamo amplification on their own, they do organize the magnetic field along the spiral density arms. 

The right panel of Fig. \ref{fig:m3b10r10t400_RZCut_phi192_r4_VB} shows the structure of the magnetic field in the same meridional plane as the left panel. The toroidal component $B_\varphi$ is maximal inside the wakes. This follows from the compressive motions that simultaneously accumulate mass and magnetic flux. Omitting resistivity, one might expect the same enhancement in gas density and in toroidal magnetic field for tightly wound spirals. As both increase by a factor of $\approx 5$, we deduce that the accumulation of magnetic flux into spiral wakes proceeds on short timescales compared to the typical resistive diffusion time $\sim h^2 / \eta$ over the lengthscale of the spirals. 

The poloidal magnetic field lines drawn in Fig. \ref{fig:m3b10r10t400_RZCut_phi192_r4_VB} are mostly horizontal in the disk midplane and vertical in its corona. The vertical component is predominantly oriented toward the midplane on both sides, such that the net magnetic flux passing vertically through the disk remains approximately zero at all radii. There is therefore no net vertical magnetic flux generated by turbulence (e.g., by splitting $B_z$ into positive and negative patches), nor injection from the radial boundaries of the computational domain. As a consequence, there are no magnetic field lines connecting both vertical boundaries that could carry angular momentum vertically away or drive a magnetized wind.

\subsection{Mean vertical structure}

Because verticality --- with respect to the disk plane --- is crucial to the GI dynamo as pictured by \citet{riols19}, we take a moment to characterize the mean vertical stratification of the disk in the quasi-steady turbulent state of our reference run.

\subsubsection{Thermal stratification} \label{sec:thermo}

\begin{figure}
  \centering
  \includegraphics[width=\columnwidth]{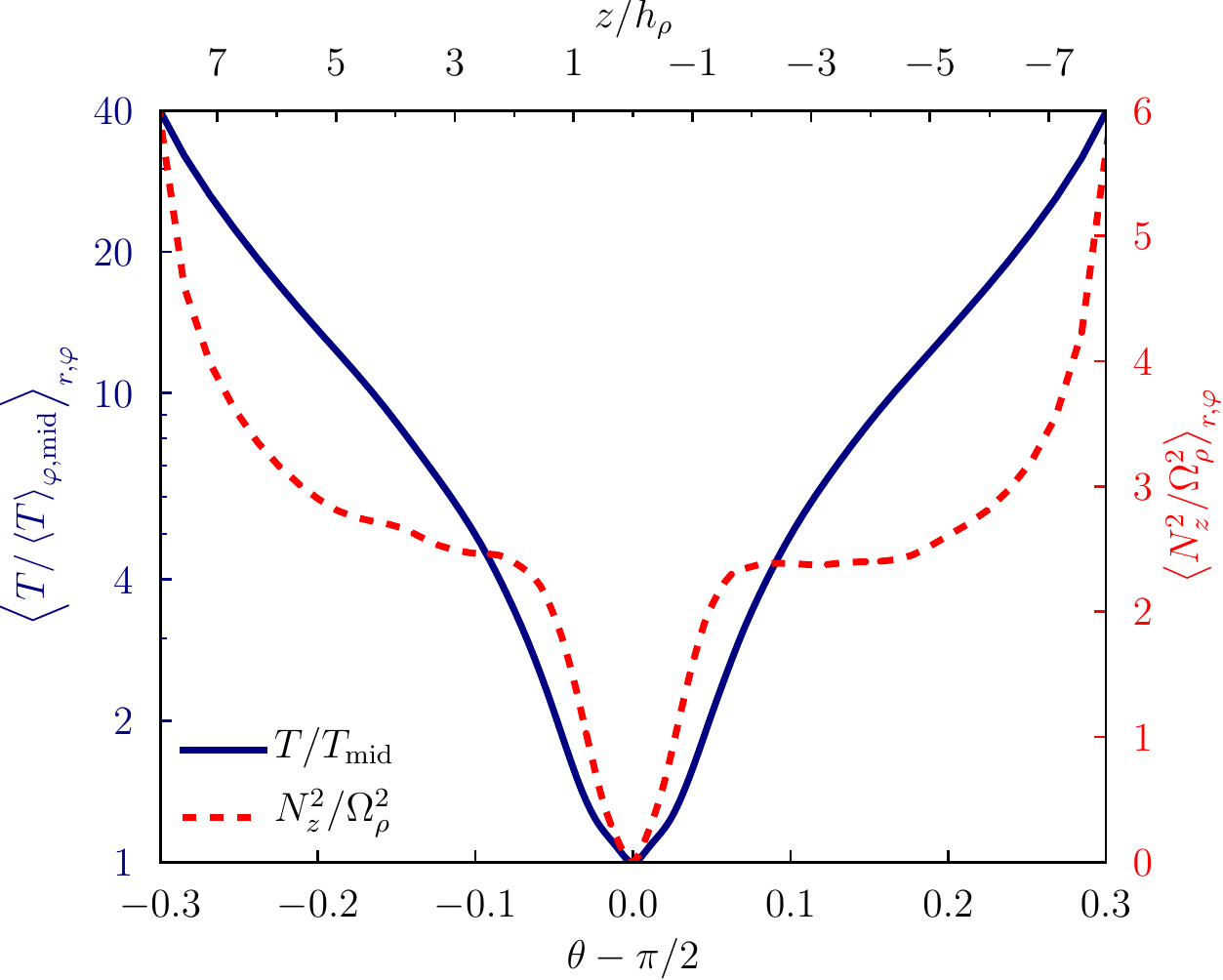}
  \caption{Thermal stratification of the disk. The solid blue line (left axis) shows the gas temperature relative to its midplane value. The dashed red line (right axis) shows the squared buoyancy frequency (Eq. \ref{eqn:bv}) relative to the local squared orbital frequency. Both profiles were averaged azimuthally, over $r/r\lin\in\left[2,8\right]$ and over time from $200t\lin$ to $400t\lin$.}
  \label{fig:m3b10r10_MFD_thermo}
\end{figure}

Anticipating the role of vertical buoyancy in the GI dynamo, and to facilitate comparison with other work, we first examine the thermal stratification of the disk. Figure \ref{fig:m3b10r10_MFD_thermo} shows that the gas temperature $T \equiv P/\rho$ increases by a factor of $\approx 40$ with height in run M3T10R10. Unlike the shearing box simulations of \citet{shi14} and \citet{riols19}, our disks are embedded in a warm corona. \citet{bethune21} attributed this feature to their choice of boundary conditions that hinder rotational support and indirectly promote pressure (thermal) support against the gravity of the central object.

To quantify the buoyant response of the gas to small vertical displacements, we define a squared buoyancy frequency
\begin{equation} \label{eqn:bv}
  N_z^2 \equiv -\frac{1}{\gamma} \left(\partial_z \Phi \right) \left(\partial_z \log\left[s\right]\right),
\end{equation}
where $s \propto P/\rho^{\gamma}$ measures the gas specific entropy, and where the gravitational potential $\Phi$ includes the contribution of the disk. This quantity is also plotted in Fig. \ref{fig:m3b10r10_MFD_thermo} after normalizing it by the squared orbital frequency and averaging. We find that $N_z^2\geq 0$ everywhere, as expected from the combination of a decreasing density and increasing temperature with height relative to the midplane. The disk is therefore convectively stable according to Schwarzschild's criterion. In fact, fitting a polytropic relation $P \propto \rho^{\Gamma}$ onto the mean vertical profiles gives $\Gamma \approx 0.7$ in run M3T10R10. This stratification is strongly sub-adiabatic, and even outside the range probed by \citet{riols18}. Based on their results, we would expect the thermal stratification to facilitate the formation of rolls near spiral wakes.

\subsubsection{Mean flows} \label{sec:meanflows}

\begin{figure}
  \centering
  \includegraphics[width=\columnwidth]{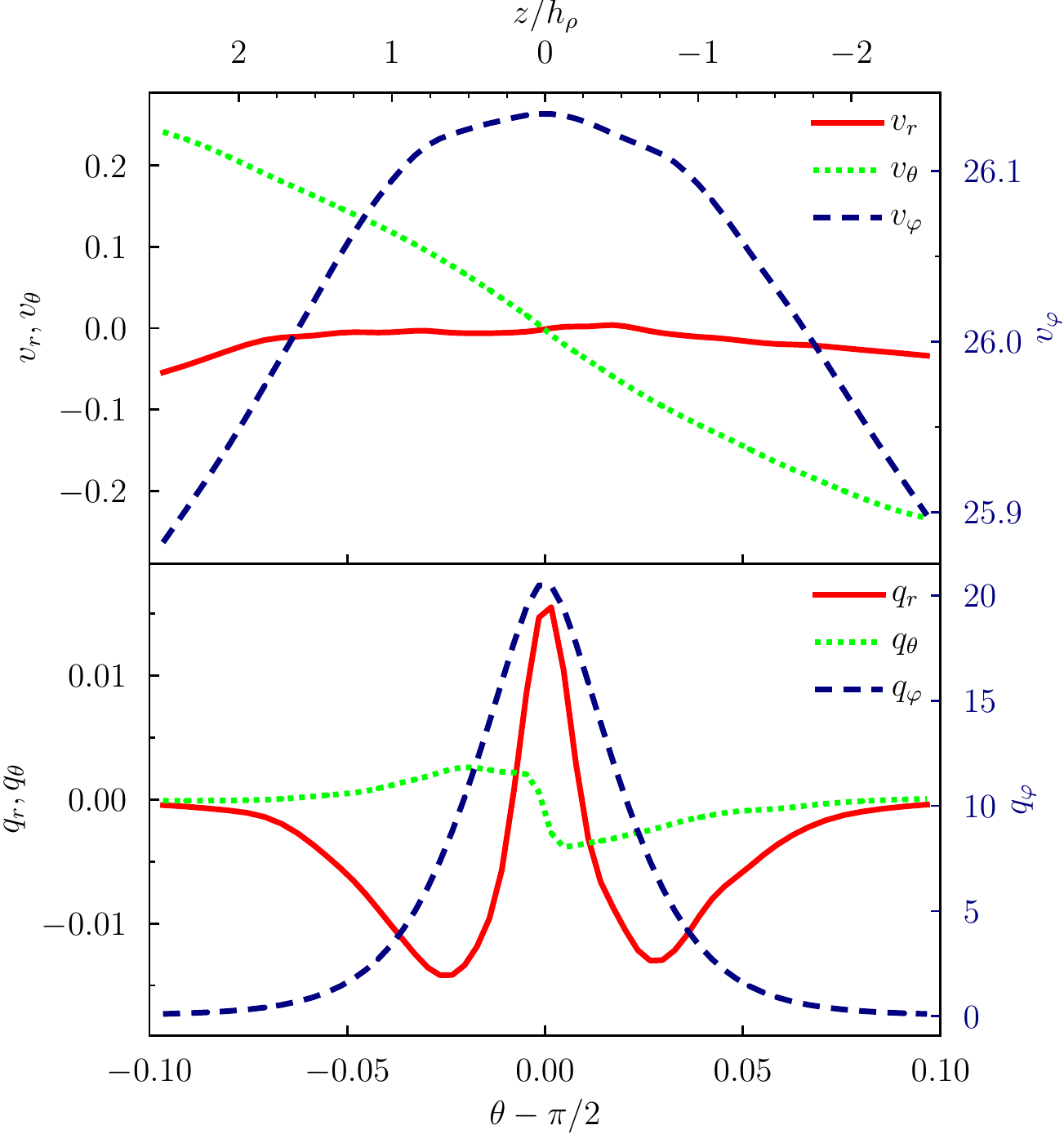}
  \caption{Meridional profiles of the mean flow as defined by Eq. \eqref{eqn:redvar}, averaged from $200t\lin$ to $400t\lin$ in run M3T10R10, and focused on the midplane region $\theta \in \uppi/2 \pm 0.1$. \emph{Upper panel}: Spherical components of the reduced gas velocity. \emph{Lower panel}: Components of the reduced mass flux. The toroidal components are measured on the right vertical axes.} 
  \label{fig:m3b10r10_EMF_mF_V}
\end{figure}

The disk also supports mean gaseous flows varying with height relative to the midplane, as we show in Fig. \ref{fig:m3b10r10_EMF_mF_V}. We distinguish the mean velocity $\bm{v}$ of the gas (upper panel), which governs the transport of magnetic flux according to Eq. \eqref{eqn:dtB}, from the mean mass flux $\bm{q}$ (lower panel) as defined by Eq. \eqref{eqn:redvar}. One may vanish while the other does not, such that mass and magnetic flux may be transported in different proportions --- even in ideal MHD.

Looking first at the radial components (solid red), the mean velocity $v_r$ (upper panel) is roughly constant and overall negative, suggesting a slow inward advection of (toroidal) magnetic flux. The mean radial mass flux $q_r$ (lower panel), on the other hand, reveals mass accretion near $\vert z/h_\rho \vert \approx 0.7$ and decretion in the midplane. We caution that the disk is not steady on viscous timescales because of our choice of initial conditions, so there is no conflict between mass decretion and outward transport of angular momentum by turbulent stresses. 

Considering the toroidal components (dashed blue, right axes), both the gas velocity and mass flux are peaked in the midplane. The orbital velocity $v_\varphi$ (upper panel) decreases with height, and its vertical shear rate $\dd V_\varphi / \dd z \approx -3\Omega_\rho$ even exceeds the radial (Keplerian) one. This vertical shear is necessary to satisfy momentum balance in conjunction with the steep thermal stratification (see Fig. \ref{fig:m3b10r10_MFD_thermo}). We do not expect it to drive a vigorous vertical shear instability because of the slow cooling timescale ($\tau>1$) and the strongly stabilizing entropy stratification \citep{urpin98, nelson13}.

Finally, the mean meridional mass flux $q_\theta$ (dotted green, lower panel) indicates that gas is accumulating in the midplane: the disk compresses vertically, possibly in order to keep $Q\sim 1$ as mass is transported radially. The meridional mass flux does eventually vanish at large $z$, assuring us that there are no appreciable mass and thermal energy exchanges between the computational domain and its boundaries. In contrast, the amplitude of the mean meridional velocity $v_\theta$ (upper panel) increases with height and reaches a plateau beyond the range shown in Fig. \ref{fig:m3b10r10_EMF_mF_V}. This is caused by correlations of the density and velocity fluctuations. On average, the high-density wakes launch slow vertical updrafts from a small area of the disk; the lofted mass then falls to the midplane through dilute but fast vertical downdrafts. These flows manifest in a nonzero mean vertical velocity upon azimuthal averaging, but with no associated net mass flux.

\subsubsection{Large-scale magnetic field}

\begin{figure}
  \centering
  \includegraphics[width=\columnwidth]{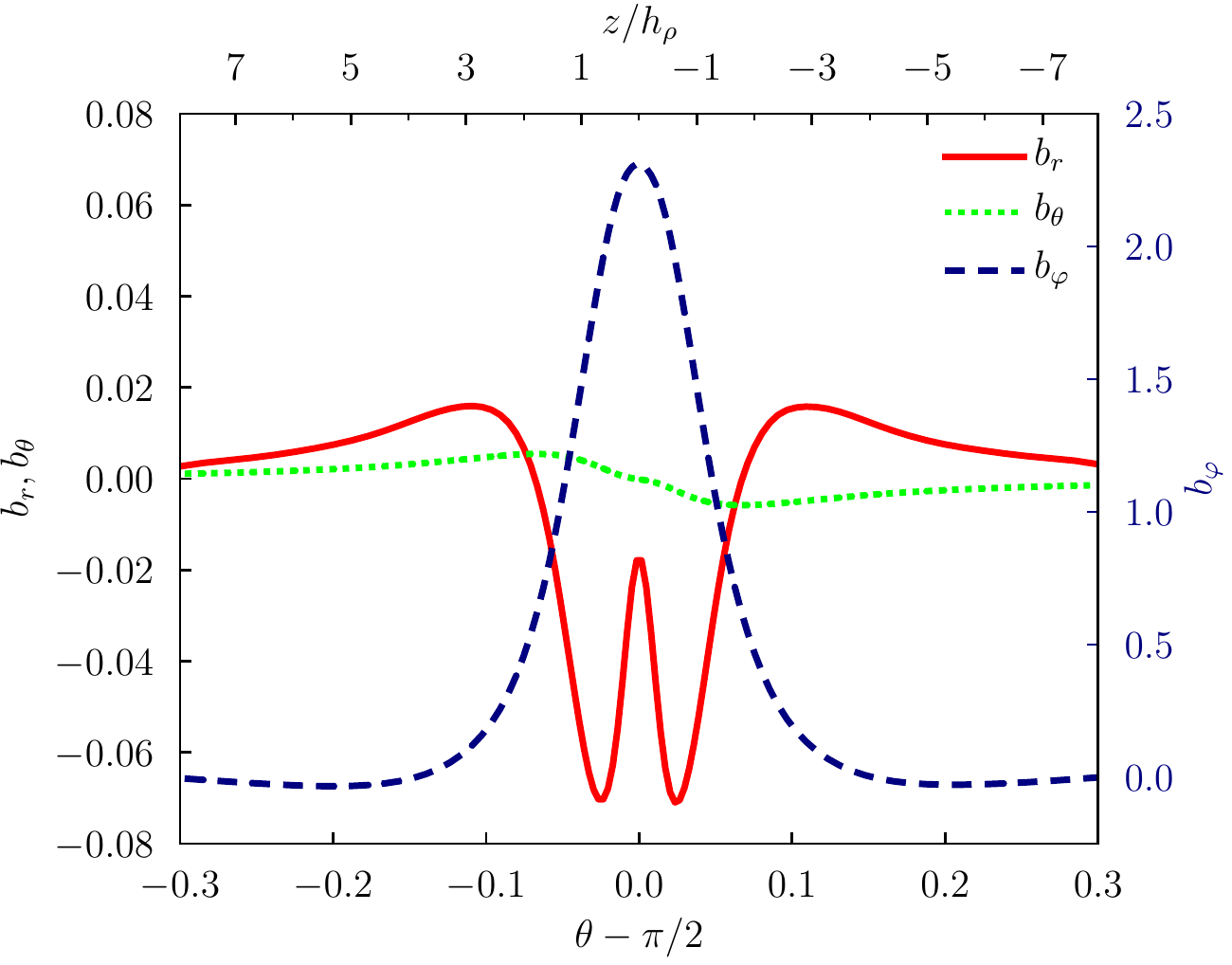}
  \caption{Meridional profiles of the three components of the reduced magnetic field as defined by Eq. \eqref{eqn:redvar} and averaged from $200t\lin$ to $400t\lin$ in run M3T10R10; the toroidal component $b_\varphi$ is measured on the right axis.}
  \label{fig:m3b10r10_EMF_B_extended}
\end{figure}

Figure \ref{fig:m3b10r10_map_b2} shows the radially global structure of the large-scale dynamo mode. In this section, we use the weighted radial average \eqref{eqn:redvar} to extract its characteristic \emph{vertical} structure. The resulting profiles are plotted in Fig. \ref{fig:m3b10r10_EMF_B_extended}. 

The radial component $b_r$ (solid red) is negative inside $\vert z/h_\rho\vert \gtrsim 1.7$ and positive further away from the midplane. It reaches at most seven per cent in amplitude, with two minima at $\vert z/h_\rho\vert \approx 0.7$. The meridional component $b_\theta$ (dotted green) is positive in the upper half of the disk ($\theta-\uppi/2<0$) and negative in the lower half, as observed in the right panel of Fig. \ref{fig:m3b10r10t400_RZCut_phi192_r4_VB}. It is clearly the weakest component, with an amplitude at most ten times smaller than $b_r$. Finally, the toroidal component $b_\varphi$ (dashed blue, right axis) is positive within $4h_\rho$ from the midplane and only slightly negative beyond, with a maximum amplitude of about $2.3$ at the midplane. We conclude that the mean magnetic field is mostly toroidal and confined inside the disk. 

According to Eq. \eqref{eqn:dtB}, the net toroidal flux $\int B_\varphi \dd S$ crossing any meridional plane (constant $\varphi$) can only change via boundary effects; any dynamo operating inside the domain would generate positive and negative flux in equal proportions. We do observe the growth of a net toroidal flux at all radii, indicating a nonzero circulation of the EMF along the boundary contour. Because we enforce $\left(B_\varphi,V_\varphi,V_\theta\right)=0$ in the ghost cells of the meridional boundaries, the ideal EMF $\mathcal{E}_r$ should approximately vanish there. The growth of a net positive toroidal flux must then be caused by the Ohmic diffusion of negative $B_\varphi<0$ contributions through the boundaries, and by any residual ideal $\mathcal{E}_r$ since it is not exactly controlled on the boundaries.

\section{Analysis of the large-scale kinematic dynamo} \label{sec:kinematic}

In this section, we examine the growth of a large-scale magnetic field during the $B^2/P \ll 1$ phase of our reference run M3T10R10. As mentioned in Sect. \ref{sec:magnetup}, the induction Eq. \eqref{eqn:dtB} then reduces to a linear initial-value problem for the magnetic field $\bm{B}$. In Sect. \ref{sec:locdyn} we use this radially averaged equation to retrace the feedback loop leading to magnetic amplification. In Sect. \ref{sec:global} we insert the (radially) local dynamo process into an idealized global disk model to understand the observed large-scale growth behaviors. 

\subsection{The GI dynamo as a radially local process} \label{sec:locdyn}

We first focus on local aspects of the GI dynamo to make contact with the shearing box simulations of \citet{riols19}. We decompose the induction Eq. \eqref{eqn:dtB} into elementary terms and then track how each component of the magnetic field acts on the other components to produce an unstable induction cycle. 

\subsubsection{Mean magnetic field induction} \label{sec:meanfieldinduc}

The curl of the ideal EMF appearing in Eq. \eqref{eqn:dtB} is equivalent to the divergence of a tensor $\bm{V}\otimes\bm{B}-\bm{B}\otimes\bm{V}$, leading to the following conservative form in cartesian coordinates:
\begin{equation} \label{eqn:dtbi}
  \partial_t B_i = \sum_j \underbrace{- \partial_j \left[ V_j B_i \right]}_\text{$A_{ji}$} + \sum_j \underbrace{\partial_j \left[ B_j V_i \right]}_\text{$S_{ji}$} \,+\,\mathrm{Ohm}.
\end{equation}
The component $A_{ji}$ represents the transport of $B_i$ along the velocity $V_j$, including advection and compression. The terms $S_{ji}$ represent the stretching of components $B_j$ into $B_i$ by velocity gradients. The diagonal terms $A_{ii}$ and $S_{ii}$ cancel each other as $B_i$ is only sensitive to transverse velocity components $V_{j\neq i}$. Using spherical coordinates, we compute the curl of the ideal EMF with the above notation but including the appropriate geometrical factors inside the derivatives. For example, $A_{r\varphi}\equiv -\left(1/r\right)\partial_r\left[r V_r B_\varphi\right]$ denotes the radial transport of $B_\varphi$.

Next, we decompose any flow variable $X$ as an axisymmetric part $\brac{X}_{\varphi}$ plus a fluctuating part $X^{\prime}$. After azimuthal averaging, the ideal EMF becomes
\begin{equation} \label{eqn:meanfluc}
  \brac{\bm{\mathcal{E}}}_{\varphi} = -\brac{\bm{V}}_{\varphi} \times \brac{\bm{B}}_{\varphi} -\brac{\bm{V}^{\prime} \times \bm{B}^{\prime}}_{\varphi}.
\end{equation}
We denote by $A_{ji}^{\prime}$ and $S_{ji}^{\prime}$ the transport and stretching terms appearing in the azimuthally averaged Eq. \eqref{eqn:dtbi} --- affecting the mean field $\brac{\bm{B}}_\varphi$ --- and arising from the rightmost term of Eq. \eqref{eqn:meanfluc}, that is, the correlation of the fluctuations $\bm{V}^\prime$ and $\bm{B}^\prime$.

To isolate the large-scale component of the magnetic field, we also average Eq. \eqref{eqn:dtbi} in the radial dimension after normalizing it by $\Omega_\rho \bar{B}$ at every radius. As for the other reduced variables, defined in Eq. \eqref{eqn:redvar}, this procedure effectively flattens the global gradients in the disk and allows different radii to contribute at the same level to the average.

\subsubsection{Mean toroidal magnetic flux} \label{sec:dtbp}

\begin{figure}
  \centering
  \includegraphics[width=\columnwidth]{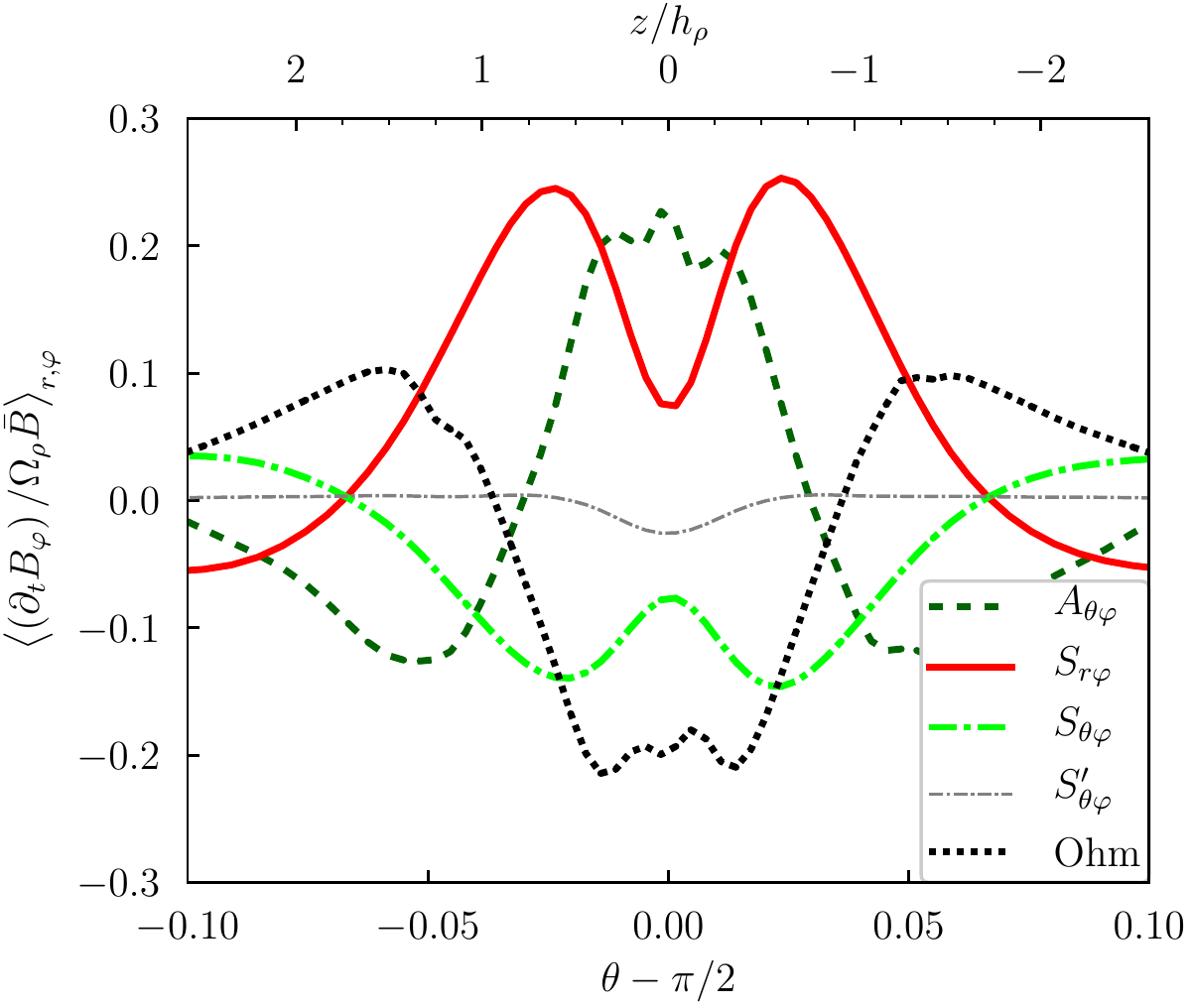}
  \caption{Contributions to the induction of a mean toroidal field after normalizing Eq. \eqref{eqn:dtbi} by $\Omega_\rho \bar{B}$ and averaging azimuthally, radially over $r/r\lin \in \left[2,8\right]$, and over time from $200t\lin$ to $400t\lin$ in run M3T10R10.}
  \label{fig:m3b10r10_EMF_DTBT_dtbp}
\end{figure}

\begin{figure}
  \centering
  \includegraphics[width=\columnwidth]{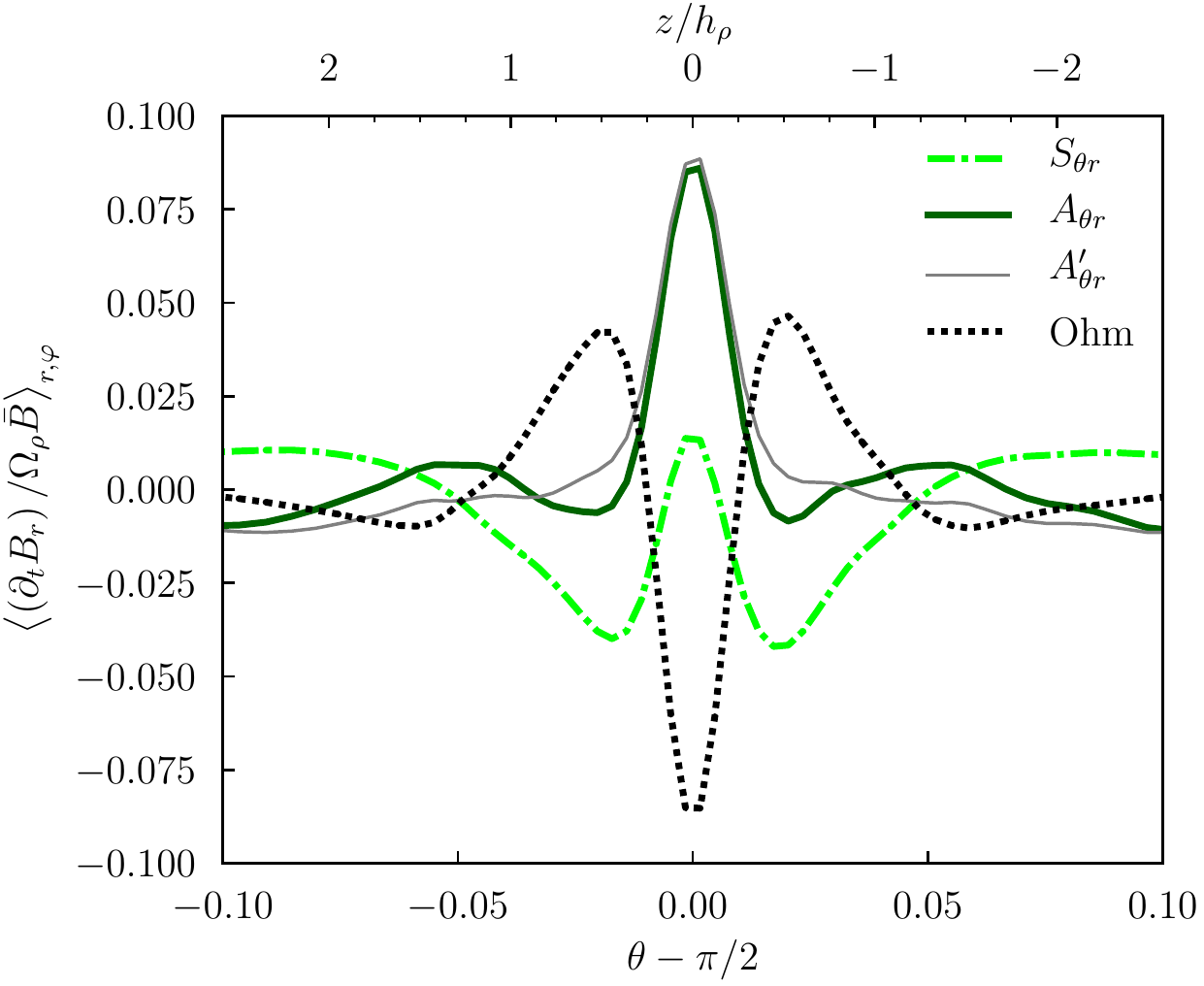}
  \caption{Same as Fig. \ref{fig:m3b10r10_EMF_DTBT_dtbp} but for the induction of a mean radial field.}
  \label{fig:m3b10r10_EMF_DTBT_dtbr}
\end{figure}

Because the toroidal component $B_\varphi$ is dominant, we take it as the starting point of a feedback loop. The different contributions to the induction of a mean $B_\varphi$ are represented in Fig. \ref{fig:m3b10r10_EMF_DTBT_dtbp}. The toroidal component is controlled by the largest number of terms, to which we can nevertheless associate simple interpretations. 

The $A_{\theta\varphi}$ term (dashed, dark green) represents the accumulation of $B_\varphi$ in the midplane due to the meridional velocity. It is dominated by the action of the converging mean flow discussed in Sect. \ref{sec:meanflows} ($v_\theta$ in Fig. \ref{fig:m3b10r10_EMF_mF_V}). The Ohmic contribution (dotted black) opposes $A_{\theta\varphi}$, working to reduce $B_\varphi$ in the midplane by spreading it vertically. The symmetry of these two terms translates into an approximate advection-diffusion balance in the vertical direction. 

The other term acting to induce a positive $B_\varphi$ near the midplane is $S_{r\varphi}\equiv \left(1/r\right) \partial_r \left[r V_\varphi B_r\right]$ (solid red), corresponding to the stretching of $B_r$ into $B_\varphi$ by the differential rotation of the disk. Its fluctuating part $S_{r\varphi}^{\prime}$ in fact reduces to zero, so the growth of a net $B_\varphi>0$ essentially comes from the radial shear of a net $B_r<0$. This is usually refered to as the $\Omega$ effect in classic mean-field dynamo theory, and it is clearly operating in our simulations, in agreement with \citet{riols18a,riols19}. 

Finally, the $S_{\theta\varphi}$ term (dot-dashed, light green) is approximately mirror symmetric with $S_{r\varphi}$, albeit with a smaller amplitude. It represents the stretching of a vertical field ($B_z$) into a toroidal one ($B_\varphi$) by the gradient $\partial_z \left[B_z V_\varphi\right]$. Its fluctuation part $S_{\theta\varphi}^\prime$ is comparatively small, so $S_{\theta\varphi}$ also acts on the mean fields as an $\Omega$ effect. However, it works against the induction of a positive $B_\varphi$ and therefore against the GI dynamo. This term was absent from local simulations because they do not capture the global gradients responsible for a vertical shear in the orbital motion.

\subsubsection{Mean radial magnetic flux} \label{sec:dtbr}

We have shown that the mean $B_\varphi>0$ appears to be generated primarily by the shear of a mean $B_r<0$; we now look for the origin of this mean radial field. To do so, we plot the various terms of the averaged induction equation for $\brac{B_r}_\varphi$ in Fig. \ref{fig:m3b10r10_EMF_DTBT_dtbr}. 

The $A_{\theta r}$ term (solid, dark green) represents the vertical transport of $B_r$ by $V_\theta$. Its peak in the midplane is essentially due to the fluctuating part $A_{\theta r}^{\prime}$ (thin gray). Being opposed to the Ohmic term (dotted black), the turbulent velocity fluctuations appear to play an antidiffusive role on $B_r$ in the vertical direction, with an approximate balance similar to that found for $B_\varphi$ in Fig. \ref{fig:m3b10r10_EMF_DTBT_dtbp}.

The only term inducing $\brac{B_r}_\varphi<0$ on both sides of the midplane, and therefore enhancing the structure shown in Fig. \ref{fig:m3b10r10_EMF_B_extended}, is $S_{\theta r} \equiv \left(1/r\sin\left(\theta\right)\right) \partial_\theta \left[\sin\left(\theta\right) V_r B_\theta \right]$ (dot-dashed, light green). This term corresponds to the stretching of a vertical field ($B_z$) into a radial one ($B_R$) by a vertical gradient $\partial_z \left[ B_z V_R\right]$. It is dominated by its fluctuating part $S_{\theta r}^{\prime}$, and hence the induction of a mean $\brac{B_r}_\varphi$ cannot be directly traced back to another mean-field component. In Sect. \ref{sec:mfd} we nevertheless formulate a closure for the dynamo loop that relies solely on the mean fields.

\subsubsection{Fluctuations of the meridional magnetic field} \label{sec:dtb2}

\begin{figure}
  \centering
  \includegraphics[width=\columnwidth]{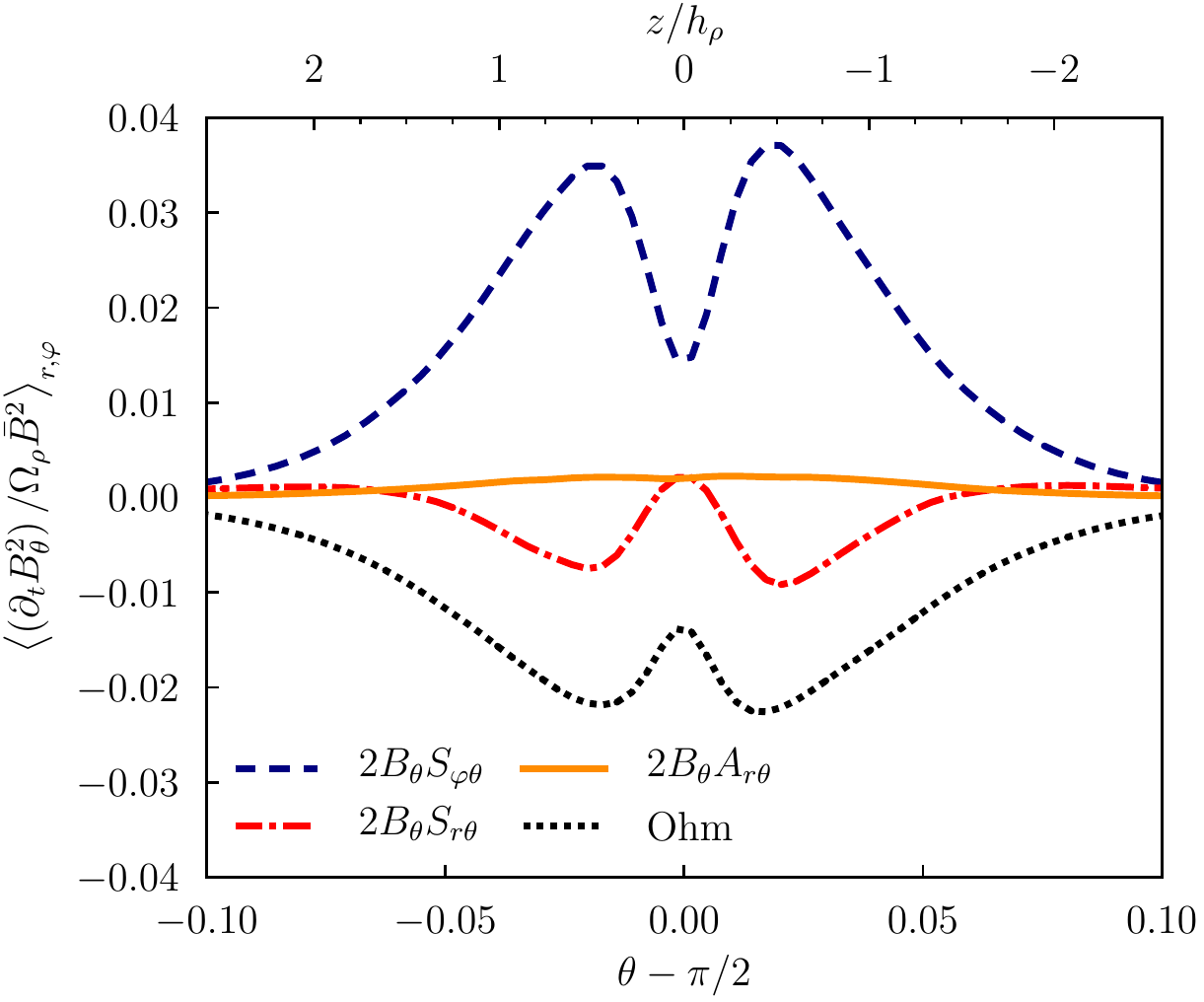}
  \caption{Contributions to the energy stored in $B_\theta^2$ after normalizing $\partial_t B_\theta^2$ by $\Omega_\rho \bar{B}^2$ at every radius and averaging azimuthally, radially over $r/r\lin \in \left[2,8\right]$, and over time from $200t\lin$ to $400t\lin$ in run M3T10R10.}
  \label{fig:m3b10r10_EMF_DTBT_dtbt2}
\end{figure}

Although the mean meridional component $\brac{B_\theta}_{\varphi}$ plays no significant role in the induction of $\brac{B_r}_\varphi$, the key term $S_{\theta r}^\prime$ does rely on nonaxisymmetric fluctuations $B_\theta^\prime$. Because $\brac{B_\theta}$ is so small, $\brac{B_\theta^2}$ essentially measures the energy in these fluctuations. Our final step is to connect this magnetic energy back to the mean fields. To this purpose, we multiply each component $A_{j\theta}$ and $S_{j\theta}$ of the induction Eq. \eqref{eqn:dtbi} by $2 B_\theta$ to obtain (twice) the rate of change of the magnetic energy, that is, $\partial_t B_\theta^2$. We normalize each term by the local $\Omega_\rho \bar{B}^2$, average them as previously, and plot them in Fig. \ref{fig:m3b10r10_EMF_DTBT_dtbt2}. 

Only the $S_{\varphi\theta}$ term (dashed blue) is positive throughout and therefore injects energy into the fluctuations of $B_\theta$. It corresponds to the generation of a vertical field ($B_z$) from a toroidal one ($B_\varphi$) by the azimuthal variations of the vertical velocity ($\partial_\varphi V_z$). This term closes the loop by generating $B_\theta^\prime$ fluctuations from the mean $B_\varphi$. These fluctuations can then induce a mean radial field, which then shears back into a mean azimuthal field.

Considering the remaining terms, $A_{r\theta}$ (solid orange) corresponds to the radial transport (advection and compression) of $B_\theta^2$ and should eventually vanish upon averaging. The $S_{r\theta}$ term (dot-dashed red) is negative inside the disk, meaning that the stretching of a radial field ($B_R$) into a vertical one ($B_z$) by radial gradients $\partial_R \left[B_R V_z\right]$ happens at the expense of magnetic energy.

The energy exchanges of the other components $B_r^2$ and $B_\varphi^2$ are also confined inside $\vert z/h_\rho\vert \lesssim 2$, confirming that the growth of the disk magnetization follows from a turbulent dynamo inside the disk, with no significant energy input from the domain boundaries. Magnetic energy is mainly injected into $B_\varphi^2$ but also slightly into $B_r^2$, while Ohmic resistivity dissipates most of it into heat and brings the disk close to marginal dynamo stability.

\subsubsection{Illustration of the local dynamo process} \label{sec:dynamorecap}

\begin{figure}
  \centering
  \includegraphics[width=\columnwidth]{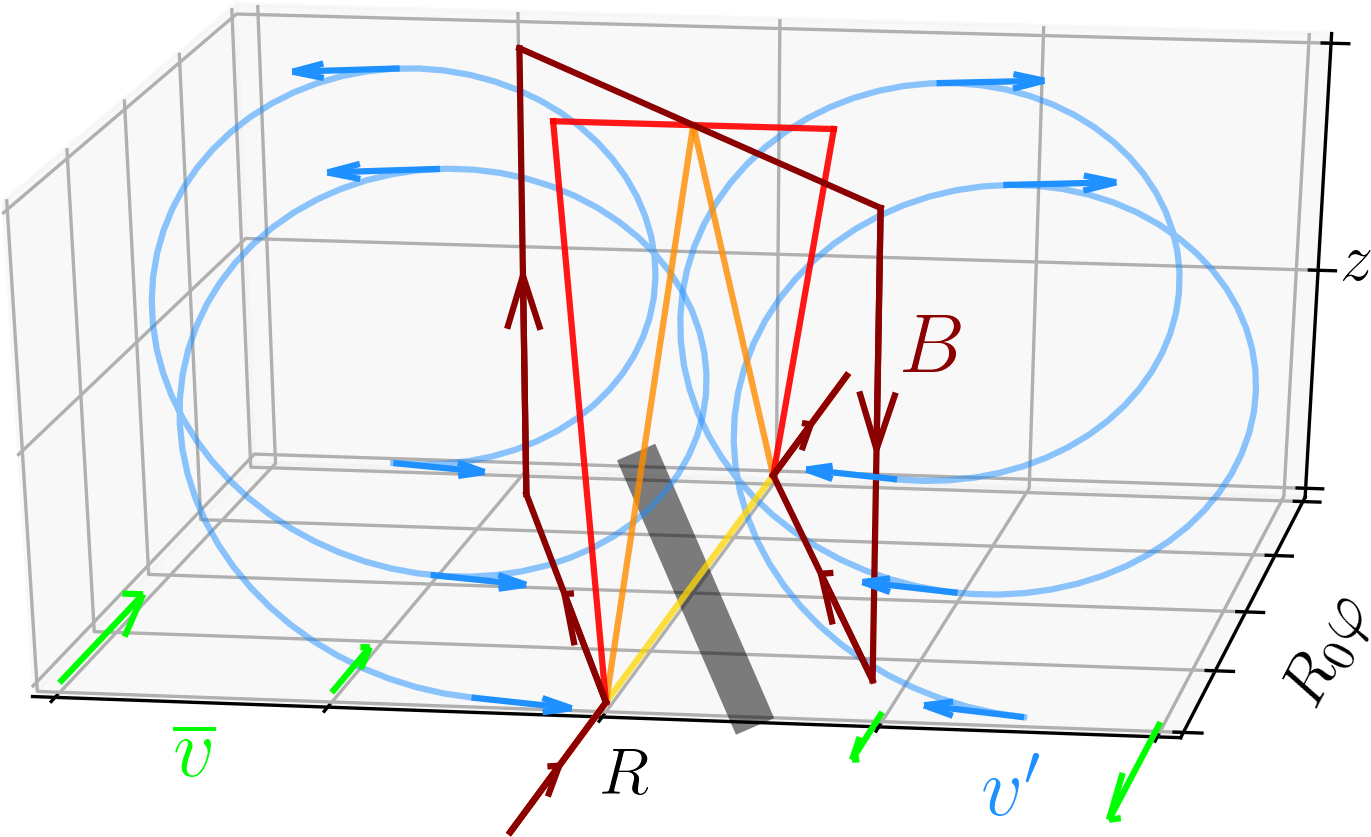}
  \caption{Dynamo mechanism as seen in the frame of a spiral wake; see the text in Sect. \ref{sec:dynamorecap} for a step-by-step description.}
  \label{fig:dynamo}
\end{figure}

We have recovered the ingredients of the GI dynamo identified by \citet{riols19} after azimuthal and radial averaging of our global simulation results. We now provide a more geometric interpretation of its successive steps in Fig. \ref{fig:dynamo}. As noted in Sect. \ref{sec:hydroturb}, we could not easily identify such regular flow patterns in our simulations, so Fig. \ref{fig:dynamo} should be taken as idealized. In particular, we omit the mean vertical shear $\partial_z V_\varphi$ and its ability to stretch $B_z$ into $B_\varphi$ (the $S_{\theta \varphi}$ term in Fig. \ref{fig:m3b10r10_EMF_DTBT_dtbp}) in this already cramped representation. Finally, the role of the mean vertical advection of magnetic field is also neglected, even though later in Sect. \ref{sec:mfd} we suggest that it is potentially important.

In the figure, the thick gray line represents a spiral wake, inclined with respect to the azimuthal ($\varphi$) direction. The green arrows at the bottom represent the mean shear flow, while the blue arrows and helices describe nonaxisymmetric velocity fluctuations. The segmented lines represent a magnetic flux tube at four consecutive times. Initially toroidal (yellow), it is first pinched vertically away from the midplane (orange), then twisted into a radial component by the helices (light red), stretched azimuthally by the mean shear flow (dark red), and finally folded to form a loop with two flux tubes bunched in the midplane, reinforcing the azimuthal field we started with.

\subsection{Global aspects of the dynamo} \label{sec:global}

By using radially averaged quantities, we lost information concerning the radially global properties of the dynamo, as manifested in Fig. \ref{fig:m3b10r10_map_b2}. We now examine how magnetic amplification proceeds across the radial span of the disk. We propose a heuristic model to describe global dynamo modes in Sect. \ref{sec:toyfit} and apply it to estimate turbulent parameters in Sect. \ref{sec:globeig}.

\subsubsection{Heuristic model of a global dynamo} \label{sec:toyfit}

Whatever the equation(s) governing the evolution of $\bar{B}\left(R,t\right)$, the fact that $\partial_t \bar{B} \simeq \omega \bar{B}$ with a constant $\omega$ in Fig. \ref{fig:m3b10r10_rad_growth} means that we are observing a global eigenmode. It is instructive to build a physically motivated model for the evolution of $\bar{B}$ that could connect the local and global physics of the problem. 

\citet{bethune21} argued that the properties of GI turbulence were essentially local within this simulation setup. We also know from the simulations of \citet{riols19} and from our analysis of Sect. \ref{sec:locdyn} that the GI dynamo can be explained using only radially local quantities. It therefore seems reasonable to assume that magnetic amplification is driven by a local source term $\varpi \Omega \bar{B}$ operating on local dynamical times, where the dimensionless factor $\varpi$ encapsulates the details of the local dynamo efficiency (i.e., its growth rate). 

The second major actor is radial magnetic diffusion, as provided by Ohmic resistivity and turbulence, which provides the ``connective tissue'' between different radii and thus permits the development of coherent global modes. Our setup is designed such that dimensionless numbers constructed from local quantities --- the aspect ratio $h/R$, cooling timescale $\tau$, Ohmic Reynolds number $\Rm$ --- are independent of radius. We may as well expect the turbulent Reynolds number to be radially constant. The total magnetic diffusivity would then take the form $\zeta \Omega h^2$, where $\zeta$ denotes an effective magnetic Reynolds number. We therefore propose a simple reaction-diffusion equation to describe the evolution of the magnetic field amplitude:
\begin{equation} \label{eqn:toyfit}
  \frac{\partial \bar{B}}{\partial t} \simeq \varpi \Omega \bar{B} + \frac{1}{R}\frac{\partial}{\partial R} \left( R \zeta \Omega h^2 \frac{\partial \bar{B}}{\partial R}\right),
\end{equation}
consisting of local growth plus radial diffusion. 

In the next paragraphs, we solve Eq. \eqref{eqn:toyfit} numerically. But assuming that $\bar{B}\propto \text{e}^{\omega t}$, and after a change of variable, we can manipulate it into a Schr\"odinger form (not shown) from which we extract a mode's turning point $R_{\rm{tp}}$ --- delimiting the region $R<R_{\rm{tp}}$ where a dynamo mode is localized. An approximate equation for $R_{\rm{tp}}$ is $\omega \simeq \varpi \Omega(R_{\rm{tp}})$, which reveals that the global growth rate of the dynamo mode $\omega$ is set by the local forcing rate at the mode's outermost (and hence slowest) part\footnote{The growth of the global MRI is limited in an analogous way, though in that context, it is magnetic tension rather than diffusion that links disparate radii into a coherent global mode \citep{LFF15}.}.

\subsubsection{Fitting global eigenmodes} \label{sec:globeig}

\begin{figure}
  \centering
  \includegraphics[width=\columnwidth]{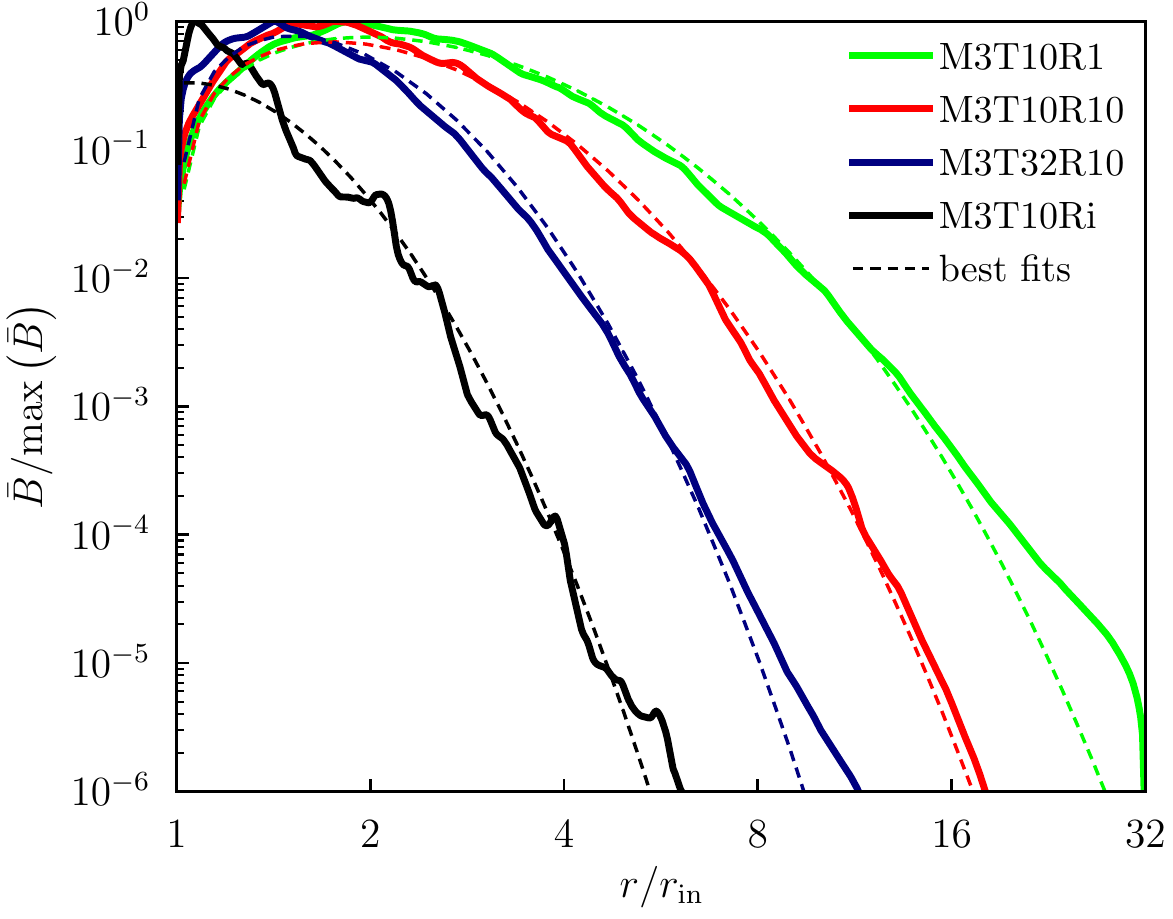}
  \caption{Normalized profiles of the magnetic field amplitude $\bar{B}$ measured before dynamo saturation in four simulations featuring a steady exponential growth (solid lines; see legend), and the best-fit eigenmodes from Eq. \eqref{eqn:toyfit} having the same global growth rate (dashed lines).}
  \label{fig:toyfits}
\end{figure}

We validated Eq. \eqref{eqn:toyfit} a posteriori by comparing its eigenmodes to simulation data, namely the radial profile of $\bar{B}$ at the time that the average disk magnetization reaches $B^2/P=10^{-2}$ anywhere. We computed eigenmodes of \eqref{eqn:toyfit} for given pairs of $\left(\varpi,\zeta\right)$ by successively integrating it in time and rescaling the solution until convergence, imposing $\bar{B}=0$ at both boundaries\footnote{Except in run M3T10Ri for which we imposed $\partial_R \bar{B}=0$ at $R=r\lin$ only to better match simulation profiles; see Fig. \ref{fig:toyfits}.} and using the initial simulation profiles of $\bar{B}$, $\Omega$ and $h$. Noting that the eigenmodes of \eqref{eqn:toyfit} are invariant when changing $\varpi$ and $\zeta$ in the same proportions, we included the measured growth rate $\omega/\Omega\lin$ as an additional optimization constraint. We then used the uncertainty on the measured global growth rate to evaluate uncertainties on the best fit parameters $\varpi$ and $\zeta$. We consistently obtained good fits when the disk supports a steady exponential magnetic growth, as illustrated in Fig. \ref{fig:toyfits} for a subset of simulations. 

In the reference simulation M3T10R10, the optimal local growth rate $\varpi \Omega \approx 7.3\times 10^{-2} \Omega$ equals the measured global growth rate $\omega \approx 1.6\times 10^{-3} \Omega\lin$ at a radius $r/r\lin \approx 14$, which we take to be the turning point of the mode. Regarding the effective magnetic Reynolds number $\zeta \approx 2.7$, it is nearly four times smaller than the prescribed Ohmic Reynolds number $\Rm=10$. This difference is too large to be explained by variations of the disk scale height $h$ (see Sect. \ref{sec:goveq}), implying a contribution from turbulent diffusion. We come back to this idea when probing the ideal MHD limit in Sect. \ref{sec:idealimit}.

Although the best fit parameters $\varpi$ and $\zeta$ are well constrained by our simulation data, they should be interpreted with caution since Eq. \eqref{eqn:toyfit} was not derived from first principles. For example, our local growth rates $\varpi \sim 10^{-2}$ are smaller than those reported by \citet{riols19} in the same parameter regime. If a direct comparison can be made with local simulations, then this difference could result from the new global effects, such as radial transport ($\zeta$ in Eq. \ref{eqn:toyfit}) and vertical shear ($S_{\theta\varphi}$ in Fig. \ref{fig:m3b10r10_EMF_DTBT_dtbp}).

\section{Mean-field model of the GI dynamo} \label{sec:mfd}

The dynamo process summarized in Sect. \ref{sec:dynamorecap} can be understood in terms of stretching, twisting, and folding magnetic field lines. This terminology provides a satisfying geometric interpretation within classic dynamo theory \citep{moffatt19,rincon19}. It is idealized, however, as the generation of a net poloidal field out of a toroidal one depends on the statistical properties of turbulence \citep{parker55a}. As a compromise, one may posit a relationship between the turbulent EMF and the local mean fields \citep[\citealt{krause16}, \citealt{brandenburg18}, and example applications by][]{vishniac97,rekowski03,fendt18}. This approach is justified by the relatively low $\Rm$ in our simulations.

In this section, we establish such a mean-field closure for the turbulent EMF. The practical goal is to generate $\brac{B_r}_\varphi$ directly from $\brac{B_\varphi}$. The fundamental goal is to better understand the action of GI turbulence on magnetic fields. We formulate a closure ansatz in Sect. \ref{sec:mfdlaw}, constrain its parameters from simulation data in Sect. \ref{sec:mfdcoef}, exhibit the resulting mean-field equations in Sect. \ref{sec:mfdeqs}, and identify their destabilizing constituents in Sect. \ref{sec:mfd_eig}.

\subsection{Closure ansatz for the turbulent EMF} \label{sec:mfdlaw}

We seek a relation between the azimuthally averaged turbulent EMF, $\brac{\bm{\mathcal{E}}^\prime}_\varphi\equiv \brac{-\bm{V}^\prime\times\bm{B}^\prime}_\varphi$, and the mean (axisymmetric) field $\brac{\bm{B}}_\varphi$. The simplest closure is a proportionality $\bm{\mathcal{E}}^\prime \propto \bm{B}$, allowing magnetic amplification when coupled to a mean shear --- the so-called $\alpha$-$\Omega$ dynamo. It has been suggested that the GI dynamo may belong to this family \citep{riols19,riols21}, and we tested this hypothesis against our simulations. 

We normalized $\bm{\mathcal{E}}^\prime$ by its characteristic scale $\bar{V} \bar{B}$ and averaged it over $\left(r,\varphi\right)$ to obtain mean meridional profiles of $\bm{\epsilon}^\prime$ as defined by Eq. \eqref{eqn:redvar}. This is similar to Sect. \ref{sec:locdyn}, where we removed any radial dependence of the induction equation and could thus extract clear profiles. The next step is to connect the reduced EMF $\bm{\epsilon}^\prime$ to dimensionless vectors derived from the mean magnetic field. This relation must be linear during the kinematic induction stage, leading to the following ansatz:
\begin{equation} \label{eqn:mfdlaw}
  \bm{\epsilon}^{\prime} = \alpha^{\prime} \cdot \bm{b} + \beta^{\prime} \cdot \bm{j},
\end{equation}
restricted for simplicity to the reduced magnetic field $\bm{b}$ and electric current density $\bm{j}$. The $\alpha^{\prime}$ matrix plays its conventional role of proportionality, whereas $\beta^{\prime}$ can be interpreted as a turbulent resistivity matrix. For simplicity, we assumed these coefficients to be constant, independent of height \citep[unlike, e.g.,][]{gressel15}. Because $\partial_t \brac{B_R}_\varphi = \partial_z \brac{\mathcal{E}_\varphi}$, the most straightforward way to induce $B_R$ out of $B_\varphi$ is via a nonzero $\alpha^{\prime}_{\varphi\varphi}$. 

\subsection{Closure coefficients from simulation data} \label{sec:mfdcoef}

We computed the profiles of $\bm{\epsilon}^\prime$, $\bm{b}$, and $\bm{j}$ in every simulation featuring a steady magnetic amplification (see Sect. \ref{sec:params}). For each three component of $\bm{\epsilon}^{\prime}$, we adjusted the six optimal parameters of Eq. \eqref{eqn:mfdlaw} on the interval $\left\vert \theta - \uppi/2\right\vert \leq 0.1$ close to the midplane. We estimated these parameters in 20 independent simulation profiles evenly spaced from $200t\lin$ to $400t\lin$, and then on the corresponding time-averaged profiles. This procedure helped us identify which parameters are statistically significant: we only retained those incompatible with zero to better than three standard deviations. As a validation, we replaced the ideal EMF by the Ohmic EMF and recovered $\alpha^{\prime} \approx 0$ and $\beta^{\prime} \approx \Rm^{-1} I$, where $I$ is the identity matrix.

\begin{figure}
  \centering
  \includegraphics[width=\columnwidth]{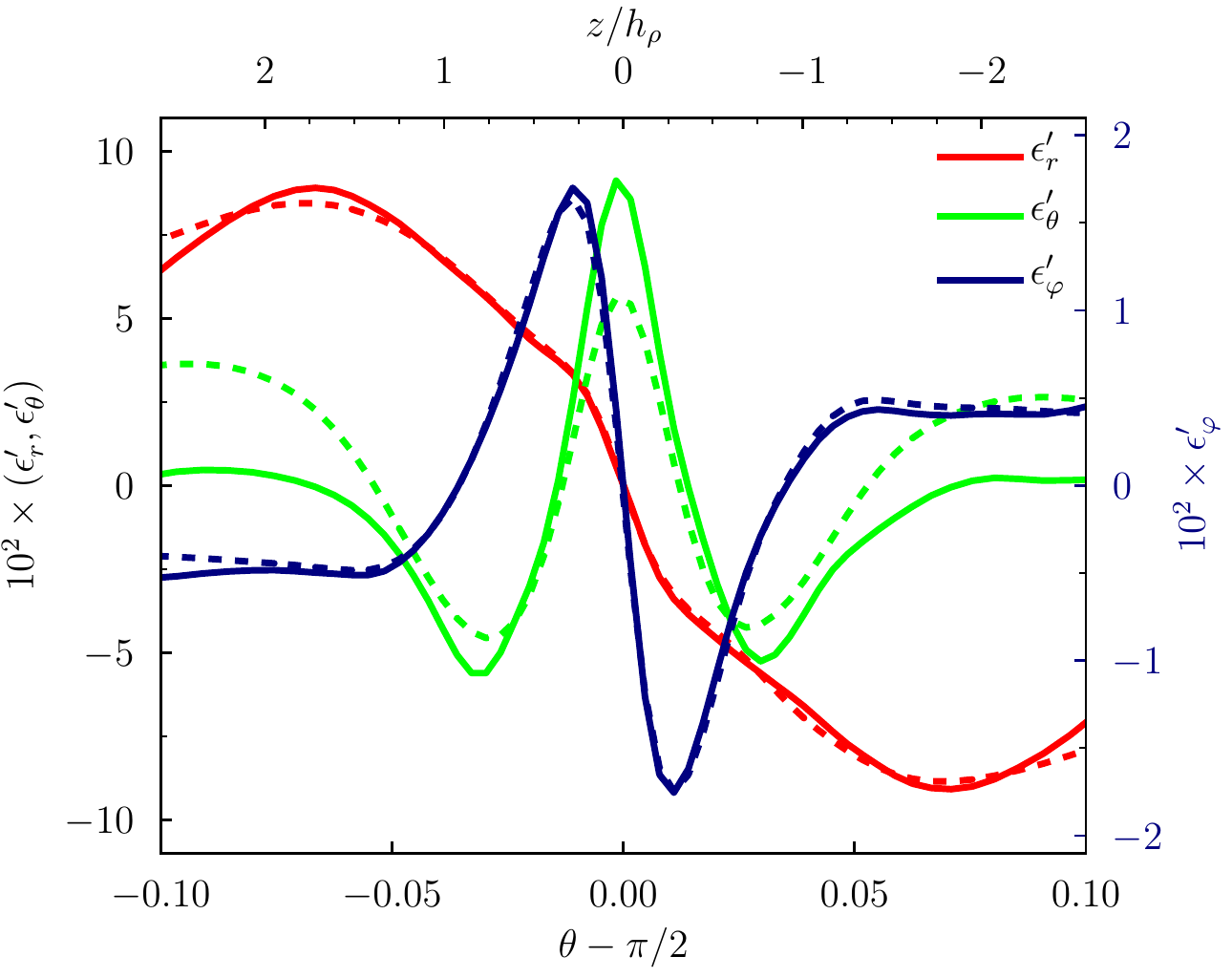}
  \caption{Meridional profiles of the reduced EMF as defined by Eq. \eqref{eqn:redvar}. The solid lines are simulation data averaged from $200t\lin$ to $400t\lin$ in run M3T10R10. The dashed lines are the best fits from Eq. \eqref{eqn:mfdlaw} over this interval. The toroidal components are measured on the right axis.}
  \label{fig:m3b10r10_MFD_E}
\end{figure}

Figure \ref{fig:m3b10r10_MFD_E} shows the time-averaged turbulent EMF in the reference run M3T10R10, along with the best fit profiles of Eq. \eqref{eqn:mfdlaw}. We could always obtain good fits for $\epsilon_r^\prime$ and $\epsilon_\varphi^\prime$ when including the $\beta^{\prime}$ matrix, but not with the $\alpha^{\prime}$ matrix alone. These matrices typically take the following form:
\begin{equation}
  \alpha^{\prime} \simeq
  \left(\begin{tabular}{ccc}
    0& $\alpha^{\prime}_{r \theta}$ & 0\\
    $\alpha^{\prime}_{\theta r}$ & 0 & 0\\
    0 & $\alpha^{\prime}_{\varphi \theta}$ & 0
  \end{tabular} \right)
  , \qquad
    \beta^{\prime} \simeq
  \left(\begin{tabular}{ccc} 
    $\beta^{\prime}_{r r}$ & 0 & $\beta^{\prime}_{r \varphi}$\\
    0 & 0 & 0\\
    $\beta^{\prime}_{\varphi r}$ & 0 & $\beta^{\prime}_{\varphi \varphi}$
  \end{tabular} \right),
  \label{eqn:emftensors}
\end{equation}
with $\alpha^{\prime}_{r \theta}>0$, $\alpha^{\prime}_{\theta r} \approx 1$, and $\alpha^{\prime}_{\varphi \theta}<0$. The diagonal coefficient $\alpha^{\prime}_{\varphi \varphi}$ is always compatible with zero, precluding the simplest link from $B_\varphi$ to $B_R$. In fact, the nonzero components of $\alpha^{\prime}$ all involve the meridional ($\theta$) dimension. Both $\alpha^{\prime}_{r\theta}$ and $\alpha^{\prime}_{\varphi\theta}$ connect to the mean meridional field $\brac{B_\theta}_\varphi$, which does not appear explicitly in the dynamo process summarized in Sect. \ref{sec:dynamorecap}. The $\alpha^{\prime}_{\theta r}$ component is the least well constrained because the quality of the fit for $\epsilon_{\theta}^\prime$ varies across simulations --- Fig. \ref{fig:m3b10r10_MFD_E} shows a rather bad case. 

In the turbulent resistivity matrix, we find $\beta^{\prime}_{\varphi \varphi}\approx -\Rm^{-1}$, which indicates a balance between Ohmic diffusion and turbulent antidiffusion of $B_r$ in the vertical direction, in agreement with Fig. \ref{fig:m3b10r10_EMF_DTBT_dtbr}. The diagonal term $\beta^{\prime}_{rr} \lesssim 0$ is always smaller than $\beta^{\prime}_{\varphi \varphi}$, implying an inefficient vertical transport of $B_\varphi$ by turbulence. The off-diagonal terms satisfy $\beta^{\prime}_{r\varphi} \approx -\Rm^{-1}$ and $\beta^{\prime}_{\varphi r} \gtrsim 0$; they rotate $B_R$ into $B_\varphi$ and vice versa, not unlike a (linearized) Hall effect. This process can destabilize axisymmetric perturbations depending on the orientation of the rotation \citep{balbus01,kunz08,squire16}, which is stabilizing in our case.

\subsection{Equations governing the mean fields} \label{sec:mfdeqs}

To lighten the notation, we use cylindrical coordinates and drop the angled brackets until Sect. \ref{sec:params}. Let $\eta^{\prime} \equiv \Omega_\rho h_\rho^2 \beta^{\prime}$ denote the scaled-up turbulent resistivity matrix. Omitting $\alpha^{\prime}_{r\theta}$ and $\alpha^{\prime}_{\varphi\theta}$ from \eqref{eqn:emftensors} and injecting the remaining coefficients into the azimuthally averaged induction Eq. \eqref{eqn:dtB} leads to the following closed system:
\begin{align}
  \partial_t B_R \simeq &-\partial_z \left[ V_z B_R\right] + \left( \eta + \eta^{\prime}_{\varphi\varphi}\right) \partial_z^2 B_R - \eta^{\prime}_{\varphi R} \partial_z^2 B_\varphi \label{eqn:mfd_dtbr},\\
  \begin{split}
    \partial_t B_\varphi \simeq &-\partial_z \left[ V_z B_\varphi\right] + \left( \eta + \eta^{\prime}_{R R}\right)\partial_z^2 B_\varphi - \eta^{\prime}_{R \varphi} \partial_z^2 B_R \label{eqn:mfd_dtbp}\\
    &+ \partial_R \left(\left[V_\varphi + \alpha^{\prime}_{zR} \bar{V}\right] B_R + \left(\eta/R\right)\partial_R\left[R B_\varphi\right]\right). 
  \end{split}
\end{align}
In both equations, the first term on the right-hand side is due to the mean vertical inflow ($v_\theta$ in Fig. \ref{fig:m3b10r10_EMF_mF_V}), the second term is vertical diffusion, and the third term is the rotation of magnetic field caused by the off-diagonal turbulent resistivity. The second line of Eq. \eqref{eqn:mfd_dtbp} consists of radial shear ($\Omega$ effect) and diffusion. 

Approximating $\bar{V} \simeq V_\varphi$ and $\alpha^{\prime}_{zR} \simeq - \alpha^{\prime}_{\theta r} \approx -1$, the radial shear term vanishes in Eq. \eqref{eqn:mfd_dtbp}. But we know from Fig. \ref{fig:m3b10r10_EMF_DTBT_dtbp} that this term ($S_{r\varphi}$) is in fact compatible with a Keplerian shear rate ($\alpha^{\prime}_{zR}=0$). This conflict comes from the radial averaging used to define $j_\theta$, leading to a degeneracy with $b_\varphi$. The price of our averaging procedure is that the entire column $\left(\beta^{\prime}_{r\theta}, \beta^{\prime}_{\theta\theta}, \beta^{\prime}_{\varphi\theta}\right) \approx 0$, and hence we lose radial turbulent diffusion in Eq. \eqref{eqn:mfd_dtbp}. To make up for it, the best fit of $\epsilon^{\prime}_\theta$ is biased by an unphysical $\alpha^{\prime}_{\theta r} \neq 0$. We can therefore discard this only remaining $\alpha^\prime$ coefficient.

Neglecting the effective resistivity $\eta+\eta^{\prime}_{\varphi\varphi} \approx 0$ in Eq. \eqref{eqn:mfd_dtbr} and the radial gradients of $RB_R$ and $R B_\varphi$ in Eq. \eqref{eqn:mfd_dtbp}, we finally obtain a radially local system of equations:
\begin{align}
  \partial_t B_R \simeq &-\partial_z \left[ V_z B_R\right] - \eta^{\prime}_{\varphi R} \partial_z^2 B_\varphi \label{eqn:mfd2_dtbr},\\
  \begin{split}
  \partial_t B_\varphi \simeq &-\partial_z \left[ V_z B_\varphi\right] - \eta^{\prime}_{R \varphi} \partial_z^2 B_R + \left(\eta + \eta^{\prime}_{RR}\right)\partial_z^2 B_\varphi  \label{eqn:mfd2_dtbp}\\
  &+ \left(\frac{\dd \log \Omega}{\dd \log R}\right) \Omega B_R,
  \end{split}
\end{align}
whose originality relative to classic mean-field dynamo models stems from the absence of an $\alpha$ effect, the apparently stabilizing pair of off-diagonal resistivity coefficients, and the presence of a compressive mean flow \citep{rincon19}.

\subsection{Mean-field dynamo route(s)} \label{sec:mfd_eig}

\begin{figure}
  \centering
  \includegraphics[width=\columnwidth]{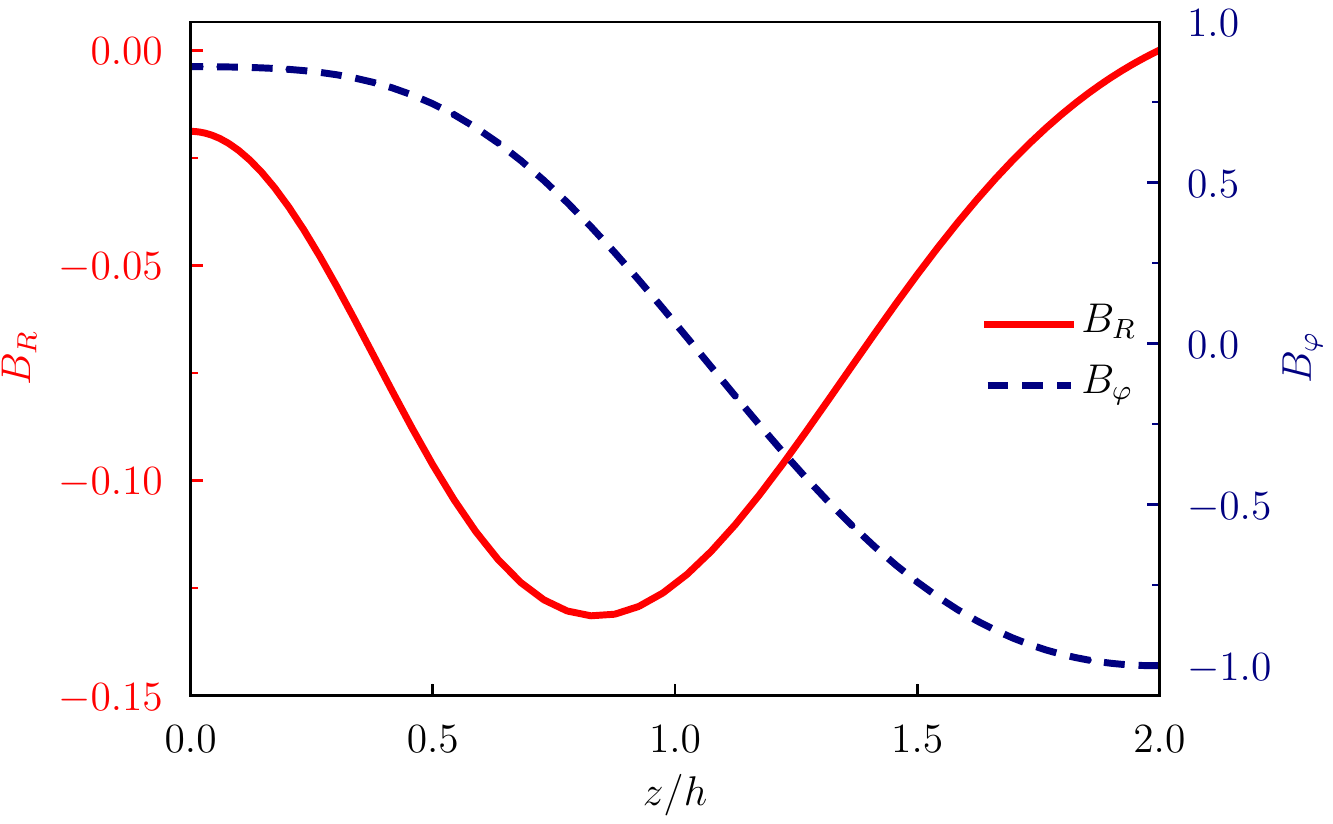}
  \caption{Eigenmode of \eqref{eqn:mfd2_dtbr}-\eqref{eqn:mfd2_dtbp} under a prescribed converging flow $V_z/\bar{V} = -0.11 z/h$, satisfying $\left(\partial_z B_R,\partial_z B_\varphi\right) = 0$ at $z/h=0$, $B_R=0$ at $z/h=2$, and $\int B_\varphi \dd z = 0$. This solution includes a Keplerian radial shear rate $\dd \log \Omega / \dd \log R = -3/2$, an Ohmic resistivity $\eta=0.1 \Omega h^2$, and the dynamo coefficients estimated in the reference simulation M3T10R10: $\beta^{\prime}_{R R}=-0.02$, $\beta^{\prime}_{R \varphi}=-0.11$, $\beta^{\prime}_{\varphi R}=7.5\times 10^{-3}$, and $\beta^{\prime}_{\varphi \varphi}=-0.1$. It is marginally unstable with a growth rate of $5.3\times 10^{-4} \Omega$.}
  \label{fig:spidy_eig}
\end{figure}

Multiplying Eq. \eqref{eqn:mfd2_dtbr} by $B_R$ and Eq. \eqref{eqn:mfd2_dtbp} by $B_\varphi$, we find two possible energy source terms that can drive instability. The first corresponds to the Maxwell stress $\propto B_R B_\varphi$ feeding on the mean orbital shear in Eq. \eqref{eqn:mfd2_dtbp}. This is the classic $\Omega$ effect identified in Sect. \ref{sec:dtbp}. The second corresponds to the vertical inflow via $\partial_t B^2 = -\left(\partial_z V_z\right) B^2$. This apparent mean flow --- resulting from correlated velocity and density fluctuations (see Sect. \ref{sec:meanflows}) --- formally permits instability even without a mean shear, and is a novelty of our reduced model.

When $V_z=0$, the off-diagonal coefficients $\eta^{\prime}_{R\varphi}$ and $\eta^{\prime}_{\varphi R}$ bring about a pair of vertically propagating waves with a retrograde rotation of the magnetic field $\left(B_R,B_\varphi\right)$. This rotation is apparent in Fig. \ref{fig:dynamo} as a toroidal field line is pinched upward and twisted from $B_\varphi>0$ into $B_R>0$. These terms operate similarly to (but are distinct from) the more common $\alpha$ effect that generates a poloidal field out of a toroidal one.

To explore the instability route associated with $V_z \neq 0$, we focus on the first two terms on the right-hand sides of Eqs. \eqref{eqn:mfd2_dtbr} \& \eqref{eqn:mfd2_dtbp}. We computed their eigenmodes as a function of $z/h \in \left[0,2\right]$ using Chebyshev polynomials over 64 Gauss-Lobatto collocation points. We sought symmetric solutions with respect to the midplane ($\partial_z B_R = \partial_z B_\varphi = 0$ at $z/h=0$), with zero net toroidal flux ($\int B_\varphi \dd z = 0$), no radial flux at the upper boundary ($B_R=0$ at $z/h=2$), and a constant compression rate $\partial_z V_z \leq 0$. These conditions are meant to represent our simulations; they do allow energy exchanges across boundary at $z/h=2$, where $\left(V_z,B_\varphi,\partial_z B_R,\partial_z B_\varphi\right)\neq 0$. We obtain unstable modes when $V_z\neq 0$ and the compression rate $-\partial_z V_z$ exceeds the waves' original frequency (see Appendix \ref{app:mfd}), indicating that the present instability need not be shear-driven. The addition of a radial shear $\dd \log \Omega / \dd \log R < 0$ in Eq. \eqref{eqn:mfd2_dtbp} further destabilizes these modes if the shear rate is small enough, and hence the $\Omega$ effect as identified in Sect. \ref{sec:dtbp} can simultaneously operate.

Finally, we injected the closure coefficients measured in run M3T10R10 into Eqs. \eqref{eqn:mfd2_dtbr} \& \eqref{eqn:mfd2_dtbp} and found the unstable eigenmode drawn in Fig. \ref{fig:spidy_eig}. Its structure qualitatively matches the corresponding simulation profiles drawn in Fig. \ref{fig:m3b10r10_EMF_B_extended}. Magnetic energy is mainly injected into $B_\varphi$, $80$ per cent of which comes from the mean vertical inflow, $17$ per cent from the Maxwell stress, and $99.7$ per cent is dissipated by Ohmic resistivity. Its growth rate of $5.3\times 10^{-4} \Omega$ makes it only marginally unstable, in line with our simulation diagnostics (see Sect. \ref{sec:dtb2}). In conclusion, the ansatz \eqref{eqn:mfdlaw} captures the mean (axisymmetric) field dynamo observed in our simulations without relying on $\alpha$ nor $\Omega$ effects. The analysis hence complicates the model constructed in Sect. \ref{sec:dynamorecap}, which, though relatively transparent, is an idealization that downplays the importance of vertical inflows. 

\begin{table*}
\caption{Parameter survey discussed in Sect. \ref{sec:params}: simulation label, initial disk mass relative to the central object, dimensionless cooling time, initial magnetic Reynolds number, total integration time, mean-field to total magnetic energy ratio \eqref{eq:meanb2}, global dynamo growth rate, local growth rate and magnetic Reynolds number from the best fit of Eq. \eqref{eqn:toyfit}. The label of the reference simulation M3T10R10 is marked with a dagger.}
\label{tab:recap}
\centering          
\begin{tabular}{l c c c c c c c c}
\hline\hline
Label & $M$ & $\tau$ & $\Rm$ & $T/t\lin$ & $\chi$ & $\omega/\Omega\lin$ & $\varpi$ & $\zeta$\\
\hline
M2T10R10 & $1/2$ & $10$ & $10$ & $1000$ & $0.60 \pm 0.05$ & $\left(2.8 \pm 0.1\right)\times 10^{-3}$ & $\left(1.34 \pm 0.04\right) \times 10^{-2}$ & $2.6 \pm 0.1$\\
M3T10Ri & $1/3$ & $10$ & $\infty$ & $850$ & $0.17 \pm 0.03$ & $\left(4 \pm 1\right)\times 10^{-3}$ & $\left(6 \pm 2\right) \times 10^{-3}$ & $11 \pm 3$ \\
M3T10R32 & $1/3$ & $10$ & $10^{3/2}$ & $1000$ & $0.38 \pm 0.07$ & $\lesssim 10^{-3}$ & $\bullet$ & $\bullet$ \\
M3T10R10$^{\dagger}$ & $1/3$ & $10$ & $10$ & $1200$ & $0.57 \pm 0.05$ & $\left(1.6 \pm 0.1\right)\times 10^{-3}$ & $\left(7.3 \pm 0.5\right) \times 10^{-3}$ & $2.7 \pm 0.2$\\
M3T10R3 & $1/3$ & $10$ & $10^{1/2}$ & $1000$ & $0.73 \pm 0.04$ & $\left(3.7 \pm 0.1 \right)\times 10^{-3}$ & $\left(1.68 \pm 0.05\right) \times 10^{-2}$ & $1.16 \pm 0.03$\\
M3T10R1 & $1/3$ & $10$ & $1$ & $500$ & $0.86 \pm 0.02$ & $\left(1.74 \pm 0.04\right)\times 10^{-3}$ & $\left(1.24 \pm 0.06\right)\times 10^{-2}$ & $0.91 \pm 0.05$\\
M3T32R10 & $1/3$ & $10^{3/2}$ & $10$ & $850$ & $0.75 \pm 0.03$ & $\left(4.9 \pm 0.4\right)\times 10^{-3}$ & $\left(1.5 \pm 0.1\right)\times 10^{-2}$ & $3.2 \pm 0.3$\\
\hline                  
\end{tabular}
\end{table*}

\section{Varying the disk parameters} \label{sec:params}

We ran simulations with different values of the disk mass $M$ relative to the central object, different dimensionless cooling times $\tau$, and different magnetic Reynolds numbers $\Rm$, as listed in Table \ref{tab:recap}. In this section, we examine how these parameters influence our findings in the kinematic regime of the GI dynamo.

One quantity that was useful in distinguishing different dynamo behaviors across our survey was the ratio of magnetic energy stored in the axisymmetric part of $\bm{B}$ relative to the total magnetic energy:
\begin{equation} \label{eq:meanb2}
\chi\left(r\right) \equiv \brac{\frac{\brac{\bm{B}}_\varphi \cdot \brac{\bm{B}}_\varphi}{\brac{\bm{B}\cdot\bm{B}}_\varphi}}_{\theta}.
\end{equation}
Smaller values of $\chi \in \left[0,1\right]$ mean that a larger fraction of magnetic energy is stored in nonaxisymmetric fluctuations, and hence this ratio measures how ordered the magnetic field is.

\subsection{Ohmic resistivity}

\citet{riols19} reported that a large-scale kinematic dynamo only occurs for moderate values of the magnetic Reynolds number (see their figure 2). At low Reynolds numbers $\Rm \lesssim 2$ the dynamo was quenched by resistivity, whereas at large numbers $\Rm \gtrsim 100$ it turned into a more stochastic and small-scale process \citep{riols18a}. We ran simulations with $\Rm$ ranging from 1 to the ``ideal'' MHD limit (see Table \ref{tab:recap}). We recall that the Ohmic resistivity $\eta$ is constant in time and varies with radius $R$ according to the prescribed value of $\Rm \equiv \Omega_{\star} h_{\mathrm{init}}^2 / \eta$, compatible with the definition used by \citet{riols19}. 

\subsubsection{High resistivity and optimal Reynolds number} \label{sec:highresi}

We ran two simulations with an increased Ohmic resistivity compared to the reference case M3T10R10, corresponding to magnetic Reynolds numbers $\Rm=\sqrt{10}\approx 3.2$ in run M3T10R3 and $\Rm=1$ in run M3T10R1. Both disks supported an exponentially growing magnetic field with global growth rates larger than in the reference run M3T10R10 (see $\omega/\Omega\lin$ in Table \ref{tab:recap}). The global growth rate is maximal in the $\Rm \approx 3.2$ case, confirming the nonmonotonic dependence of the large-scale dynamo on resistivity. In comparison, \citet{riols19} found maximal growth rates for $\Rm \approx 20$ regardless of the cooling time $\tau$. This difference of optimal $\Rm$ may point to different characteristic scales associated with the spiral wakes forming in global versus local simulations. Alternatively, it could result from the radial transport properties in the different flow geometries. 

\subsubsection{Low resistivity and oscillatory dynamo} \label{sec:lowresi}

\begin{figure}
  \centering
  \includegraphics[width=\columnwidth]{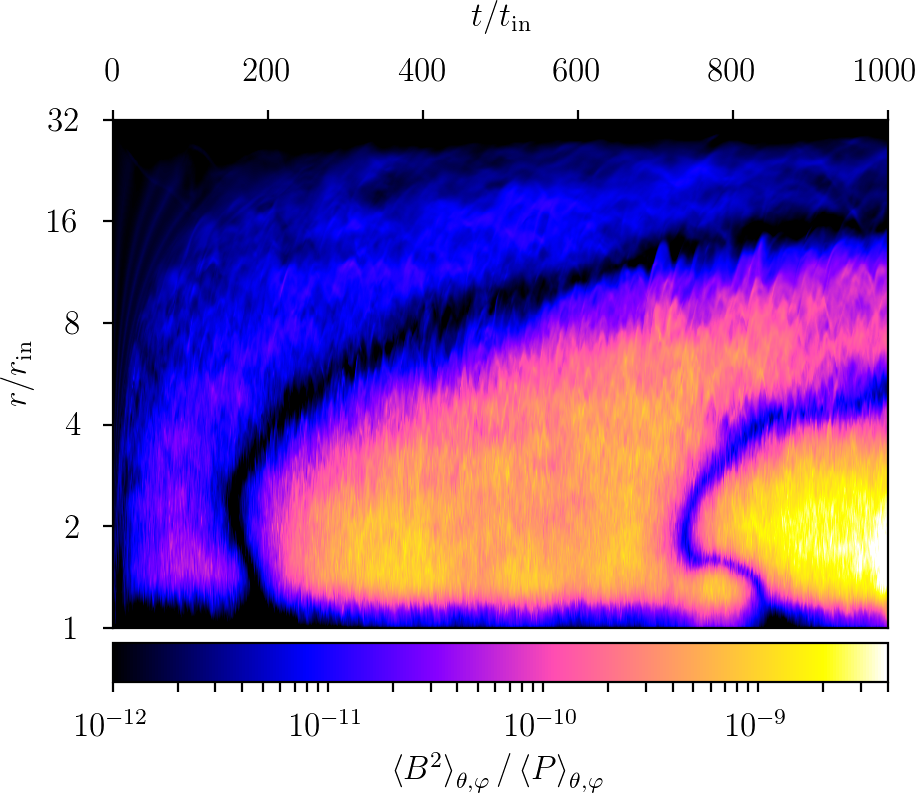}
  \caption{Evolution in time (horizontal axis) of the radial profile (vertical axis) of the disk magnetization in the low-resistivity run M3T10R32. The mean magnetic field flips sign twice while growing in amplitude.}
  \label{fig:m3b10r31_map_b2}
\end{figure}

We performed one simulation with a larger magnetic Reynolds number $\Rm=10^{3/2}\approx 32$, labeled M3T10R32. Its dynamo growth rate is smaller than in the reference case M3T10R10, following the trend found at high resistivity above. Unlike other simulations presented so far, Fig. \ref{fig:m3b10r31_map_b2} reveals that, during a time period of $1000 t\lin$, the orientation of the mean magnetic field flips sign twice while its amplitude grows. This simulation thus demonstrates the existence of an oscillatory dynamo mode that can apparently dominate on large scales. Why they were not reported by \citet{riols19} may be due in part to their initial and boundary conditions, while the oscillations reported by \citet[][see their figure 5]{deng20} were attributed to the MRI --- outside the kinematic GI dynamo regime. Whether a purely growing mode would take over on longer timescales in run M3T10R32 is unfortunately out of our computational reach.

\subsubsection{Ideal MHD limit} \label{sec:idealimit}

We ran one simulation with zero explicit Ohmic resistivity ($\Rm \rightarrow \infty$), labeled M3T10Ri in Table \ref{tab:recap}. While the GI dynamo feeds on spiral motions at the largest scales of the flow, \citet{riols19} also found a small-scale dynamo operating in the ideal MHD limit that remained sensitive to numerical resolution. Being mindful of convergence issues related to the finite numerical dissipation that dominates at the grid scale, we restrict our expectations to a qualitative transition in dynamo behavior. 

The mean-field to total magnetic energy ratio, $\chi \approx 0.17$, takes its smallest value in the ideal MHD limit, translating into a mostly disordered magnetic field. Regardless, we measured a constant (global) magnetic growth rate $\omega/\Omega\lin \approx 3.7 \times 10^{-3}$ even in this case. Surprisingly, the growth rate of the weakly resistive run M3T10R32 ($\Rm\approx 32$) is smaller than both the ideal MHD case M3T10Ri and the more resistive reference case M3T10R10. It seems unlikely that our simulations can properly resolve any small-scale dynamo in the ideal MHD limit \citep{riols18a,riols19}, so this ordering may be a fortuitous consequence of the oscillating mode prevailing in run M3T10R32.

We could also fit a global eigenmode of Eq. \eqref{eqn:toyfit} onto the profile of $\bar{B}$ in run M3T10Ri, as shown in Fig. \ref{fig:toyfits}. We thus estimate an effective magnetic Reynolds number $\zeta \approx 11$, which measures the spread of magnetic energy under the action of turbulence alone. The fact that it is only three times larger than in other runs suggests that the GI dynamo may operate in a broad range of diffusivity conditions, without relying on a large Ohmic resistivity \citep[e.g.,][]{riols21}. Remarkably, the effective Reynolds number associated with the turbulent transport of angular momentum is also $\mathcal{R}_e = 1/\alpha_{\mathrm{SS}}\approx 10$ here, where $\alpha_{\mathrm{SS}}$ is the viscosity coefficient of \citet{shakura73} determined by local thermal balance \citep{gammie01}. Their ratio gives us a crude but first estimate of the turbulent magnetic Prandtl number $\mathcal{P}_m \equiv \zeta/\mathcal{R}_e \sim 1$ in ideal gravitoturbulent disks.

\subsection{Larger initial disk mass}

We varied the disk mass by rescaling the initial gas surface density while keeping $\Sigma \propto R^{-2}$. More massive disks tend to be thicker, such that their Toomre number remains near unity after GI saturation. Run M2T10R10 has an increased disk mass $M=1/2$, corresponding to an increased aspect ratio $h_\rho / R \approx 0.06$, and it also features a steady exponential growth of magnetic energy. Its global magnetic growth rate $\omega/\Omega\lin \approx 2.8 \times 10^{-3}$ is about $1.8$ times larger than in the reference case M3T10R10. Unfortunately, the different contributions to the induction equation nearly balance each other, such that magnetic growth results from a small residual that is delicate to measure.

The meridional profiles of reduced magnetic field $\bm{b}$, as functions of $z/h_\rho$, closely resemble those of run M3T10R10 (see Fig. \ref{fig:m3b10r10_EMF_B_extended}), as does the partition of magnetic energy into $B_r^2$, $B_\theta^2$, and $B_\varphi^2$. The fractions of mean-field to total magnetic energy $\chi \approx 0.6$ are compatible in both runs. The different contributions $A_{ji}$ and $S_{ji}$ are also within 20 per cent of the reference case shown in Figs. \ref{fig:m3b10r10_EMF_DTBT_dtbp} \& \ref{fig:m3b10r10_EMF_DTBT_dtbr}. When comparing the best-fit eigenmodes of Eq. \eqref{eqn:toyfit} we find that runs M3T10R10 and M2T10R10 have similar effective Reynolds numbers $\zeta \approx 2.6$ and that the local growth rate $\varpi$ entirely accounts for the factor of $1.8$ of increase in global growth rate. The local dynamo process therefore seems to operate more efficiently in thicker (more massive) disks.

\subsection{Longer cooling time}

In steady state, the disk must generate enough heat to balance cooling, and hence the prescribed cooling time $\tau$ controls the strength of GI turbulence \citep{gammie01}. The range of cooling times accessible to global simulations is limited, however. For small values $\tau \lesssim 3$, nonmagnetized disks are prone to fragmentation \citep[e.g.,][]{lin16,booth19}. For long cooling times, the disk takes longer to reach a quasi-steady state, and turbulence may become so weak that numerical diffusion could interfere with GI saturation. For these reasons, we only considered one larger value of $\tau=10^{3/2}\approx 32$ in run M3T32R10. 

The global magnetic growth rate $\omega/\Omega\lin \approx 5\times 10^{-3}$ is approximately three times larger in run M3T32R10 compared to the reference run M3T10R10. This increased growth rate is at odds with the results of \citet[][see their figure 3]{riols19}, who found a monotonic decrease of the dynamo growth rates with increasing $\tau \in \left[ 5, 100\right]$. Unfortunately, as for the high-mass case M2T10R10 above, we could not find convincing evidence of an increased growth rate in the average induction equation. On the one hand, the turbulent velocity dispersion (hence $\bm{\epsilon}^\prime$) and the mean vertical inflow speed are smaller by a factor of $\approx 0.6$. On the other hand, the local growth rate $\varpi\approx 1.5\times 10^{-2}$ fitting an eigenmode of Eq. \eqref{eqn:toyfit} is twice larger than in the reference run.

The fact that magnetic energy grows faster at a larger $\tau\approx 32$ despite weaker velocity fluctuations is puzzling, but must stem from an increased regularity of the flow, leading to a higher yield of the dynamo process from the available kinetic energy. Although there is no hint of this in the thermal stratification of the disk --- which is nearly identical to that shown in Fig. \ref{fig:m3b10r10_MFD_thermo} --- the mean-field to total magnetic energy ratio $\chi \approx 0.75$ is indeed substantially larger in run M3T32R10. Because turbulence weakens with increasing $\tau$, we do expect an eventual decay of the magnetic growth rates for $\tau > 32$. As for the case of resistivity above, the GI dynamo may therefore admit an optimal cooling time.

\section{Saturation of the GI dynamo} \label{sec:saturated}

Except for runs in which the GI dynamo was the slowest (M3T10R32 and M3T10R1) the disk always reached a magnetized state such that $B^2/P \gtrsim 1$. We believe that the two other simulations would reach this state if integrated for longer times because a dynamo operates in both of them and only the Lorentz force $\propto B^2$ can bring it out of the linear regime. It is expected that the dynamo saturates with a plasma beta near unity because the turbulent flow is transsonic, and thus the Alfv\'enic Mach number and plasma beta are of the same order. After saturation, the strong field may support additional instabilities competing with the GI in driving turbulence, or cause the disk to puff up and possibly kill the GI, at least for some period of time.

In this section, we take a brief look at the saturated phase of the GI dynamo as it occurs in our simulations. We decompose the sources of stress in Sect. \ref{sec:satstress} and find a magnetically dominated turbulence whose accretion heating exceeds the imposed cooling. We then look at the Toomre number in Sect. \ref{sec:satq} and confirm that the GI is marginal at best, and quite likely quenched in this phase of our simulations.

\subsection{Accretion power and thermal imbalance} \label{sec:satstress}

One key quantity to study mass accretion in thin disks is the radial flux of angular momentum. We decompose it as the sum of a hydrodynamic (Reynolds) stress $\mathrm{R}_{R\varphi}$, a magnetic (Maxwell) stress $\mathrm{M}_{R\varphi}$, and a gravitational stress $\mathrm{G}_{R\varphi}$ defined by:
\begin{align} \label{eqn:stress}
  \mathrm{R}_{R\varphi} &\equiv \rho \left( v_R - \brac{v_R}_{\rho} \right) \left( v_{\varphi} - \brac{v_{\varphi}}_{\rho} \right),\\
  \mathrm{M}_{R\varphi} &\equiv - B_R B_\varphi,\\
  \mathrm{G}_{R\varphi} &\equiv \frac{\left(\partial_R \Phi_{\mathrm{disk}}\right) \left(\partial_{\varphi}\Phi_{\mathrm{disk}}\right)}{4\uppi R G}
\end{align}
\citep[see, for example,][]{balbuspap99}. To connect with standard viscous disk theory, we normalize these momentum fluxes by the vertically and azimuthally averaged gas pressure, denoting by $\alpha_{\mathrm{R,M,G}}$ the respective dimensionless coefficients, and by simply $\alpha$ the sum of all three contributions. If the conversion from orbital to thermal energy happens locally, similarly to a viscous process, then there is a unique value of $\alpha$ such that viscous heating balances the imposed cooling \citep{gammie01}. \citet{bethune21} verified that this equality holds within their framework of nonmagnetized disks, where shocks provide a heating mechanism without explicit viscosity.

\begin{figure}
  \centering
  \includegraphics[width=\columnwidth]{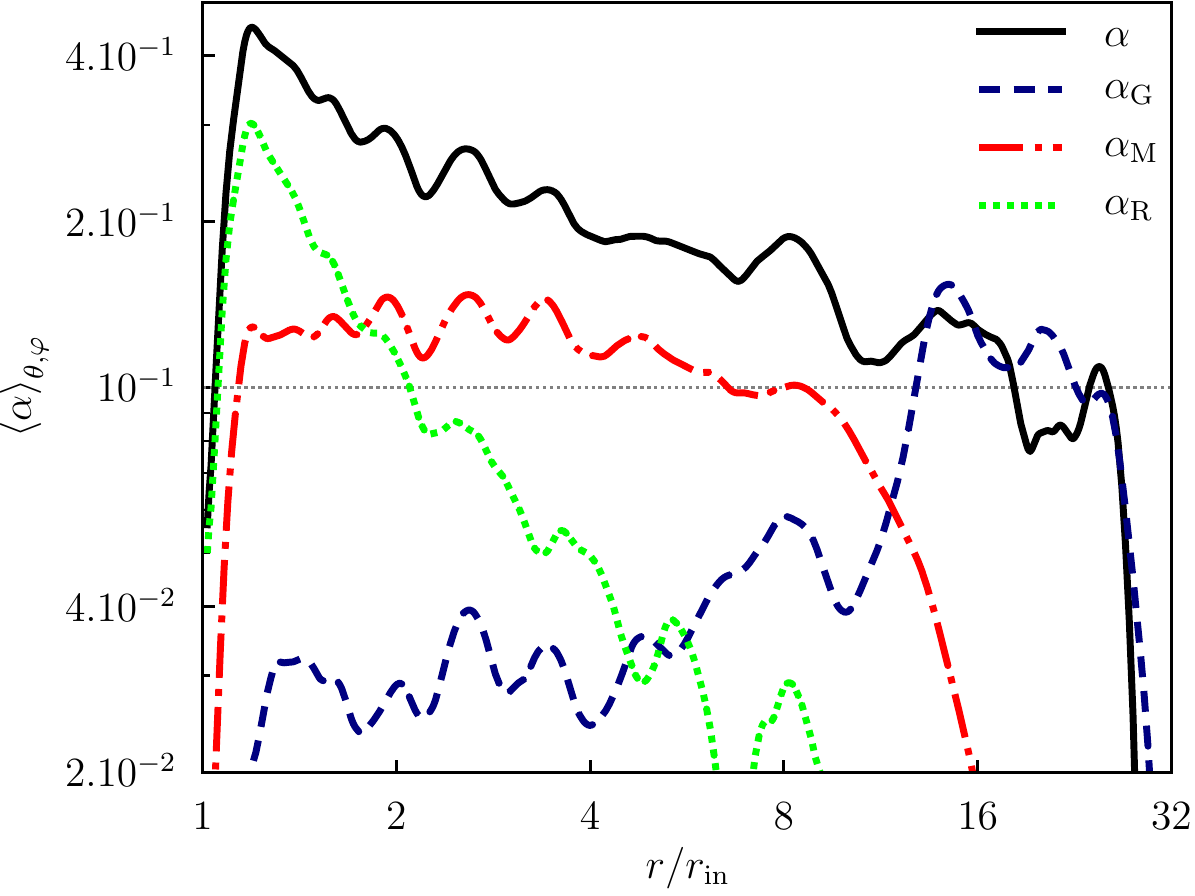}
  \caption{Radial flux of angular momentum as a function of radius after averaging in $\left(\theta,\varphi\right)$ and over time from $1136 t\lin$ to $1200t\lin$ in run M3T10R10: total stress (solid black), gravitational stress $\alpha_\mathrm{G}$ (dashed blue), magnetic stress $\alpha_\mathrm{M}$ (dot-dashed red), and hydrodynamic stress $\alpha_\mathrm{R}$ (dotted green). The thin horizontal line at $\alpha=10^{-1}$ corresponds to local thermal balance of viscous heating versus cooling.}
  \label{fig:m3b10r10_avg_alpha}
\end{figure}

We show in Fig. \ref{fig:m3b10r10_avg_alpha} the different contributions to the radial flux of angular momentum after dynamo saturation in the reference run M3T10R10. The gravitational stress $\alpha_\mathrm{G} \approx 3\times 10^{-2}$ is subdominant inside $r/r\lin \lesssim 12$, and we checked that it keeps decaying outward over time in all the dynamo-saturating runs. The magnetic stress corresponds to an effective viscosity $\alpha_{\mathrm{M}} \approx 10^{-1}$ roughly constant in the inner parts of the disk. This is the value taken by $\alpha_\mathrm{G}$ in purely hydrodynamical simulations. The Reynolds stress is maximal near the inner radial boundary, where its effective viscosity reaches $\alpha_{\mathrm{R}} \approx 3\times 10^{-1}$, but it decays rapidly with radius. \citet{deng20} reported the same trends in their ideal MHD simulations (see their figure 12).

The effective viscosity required for local thermal balance in run M3T10R10 is $\alpha=10^{-1}$. Since the total stress is substantially larger than this value, we deduce that the rate of turbulent energy injection is larger than the rate of thermal energy extraction by cooling. Whether the dissipation of turbulent energy remains a local process or not, we expect this energy excess to result in a net heating of the gas inside the computational domain. Indeed, we find that the disk thickness $h_\rho/R \lesssim 7\times10^{-2}$ overall increases as the dynamo saturates, especially in the outer parts $r/r\lin \gtrsim 4$, which implies an increase in temperature. We conclude that the disk is out of thermal equilibrium and heats up for at least $200 t\lin$ in run M3T10R10 ($300 t\lin$ in runs M2T10R10 and M3T10R3).

The disk also appears to thicken in the simulations of \citet[][comparing the central and rightmost columns of the bottom row of their figure 3]{deng20}. They argued that vertical outflows provide the additional cooling mechanism required to keep the disk steady. This is possible if there is a vertical heat flux making the disk corona hotter than its midplane --- as is our case --- such that the leaving gas has a high specific internal energy. However, our meridional boundary conditions forbid outflows and thus trap heat inside the domain. Steady thermally driven winds also require sound speeds of Keplerian amplitude at the base of the wind. This is not realized in our simulations ($c_s/v_\mathrm{K} \lesssim 0.3$ according to Fig. \ref{fig:m3b10r10_MFD_thermo}), and \citet{deng20} acknowledged that the ``clipped'' gas particles considered as outflow might actually fall back to the disk. Whether the disk can achieve a steady state after GI dynamo saturation thus remains uncertain. 

\subsection{Alternative sources of turbulence} \label{sec:satq}

\begin{figure}
  \centering
  \includegraphics[width=\columnwidth]{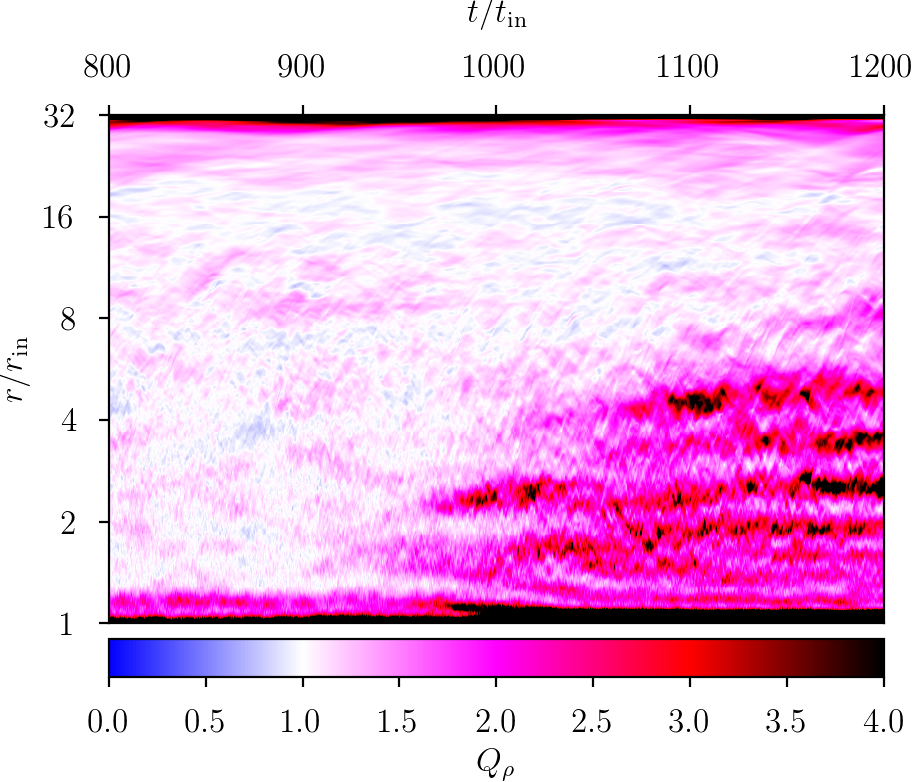}
  \caption{Evolution in time (horizontal axis) of the radial profile (vertical axis) of the density-weighted Toomre number during the dynamo saturation in run M3T10R10.}
  \label{fig:m3b10r10_map_dntom2}
\end{figure}

\citet[][]{bethune21} measured an average Toomre number
\begin{equation} \label{eqn:toomre}
  Q_\rho\equiv \frac{\kappa_\rho c_\rho}{\uppi G \Sigma} \approx 1
\end{equation}
in nonmagnetized disks, when using the density-weighted epicyclic frequency $\kappa_\rho \equiv \sqrt{ \left(2\Omega_{\rho}/R\right) \dd \left[R^2\Omega_{\rho}\right] / \dd R}$ and isothermal sound speed $c_\rho$. Figure \ref{fig:m3b10r10_map_dntom2} shows that the Toomre number $Q_\rho \gtrsim 2$ after dynamo saturation in run M3T10R10. This is partly due to the increased $c_\rho$ and partly to the decreased $\Sigma$ following radial mass transport. \citet{deng20} similarly observed $Q$ increase in their MHD disks relative to purely hydrodynamic ones (see their figure 7). Since the magnetic field has reached a thermal strength, it may also transform the linear stability of the disk with respect to GI \citep[e.g.,][]{lin14}. In the axisymmetric limit, magnetic pressure plays a stabilizing role and enhances the Toomre number by a factor of $\sqrt{1+B^2/P}$ \citep{gammie96,kim01}. The dynamo saturation appears to not only alter the gravitoturbulent flow so as to stop magnetic growth: it also appears to push the (hydrodynamic) GI toward --- and possibly beyond --- marginal stability.

As a consequence, other mechanisms may be involved in sustaining turbulence after GI dynamo saturation. The mean magnetic field remains predominantly toroidal, but it jumps from $\brac{B_\varphi}_{\theta,\varphi} \approx 0.33\bar{B}$ before saturation to $\approx 0.86\bar{B}$ after. This mean field may support MRI growth on spatially resolved wavelengths \citep{balbus92} or magnetic buoyancy may locally overcome the entropy stratification and drive Parker instabilities \citep{shu74,johansen08}. The magnetic field may also be strong enough to allow various global instabilities \citep{curry95,curry96,das18}. Evidence for an associated transition in the turbulent (and dynamo) state includes the surge of magnetic energy occurring at $t=1000 t\lin$ (see Fig. \ref{fig:m3b10r10_map_b2}). Even if GIs become inefficient according to the Toomre number, this turbulence retains enough compressible and nonaxisymmetric character to support a significant gravitational stress $\alpha_\mathrm{G}$ in Fig. \ref{fig:m3b10r10_avg_alpha}. Seeing how it is sustained for hundreds of orbits, we can exclude the on-and-off turbulent cycles suggested by \citet{riols19} over such timescales. 

Finally, we note the formation of a series of bands near $1000t\lin$ in Fig. \ref{fig:m3b10r10_map_dntom2}. They survive for at least $200t\lin$ in run M3T10R10 and also appear in the Toomre number of other simulations. They correspond to modulations of the gas surface density, with a weak and anticorrelated modulation of the gas sound speed. However, we verified that there are no genuine gas rings forming in the disk: the density fluctuations are mainly nonaxisymmetric. Such modulations did not appear in purely hydrodynamic disks, nor in the global MHD simulations of \citet{deng20}. It is unclear whether they result from the particular choices made in our numerical setup, and certainly calls for further long-term simulations outside the scope of this paper.

\section{Summary and perspectives} \label{sec:theend}

We studied the turbulent amplification of magnetic fields in gravitationally unstable disks by means of global 3D MHD simulations. Our setup was the same as that of \citet{bethune21}, but seeded with a weak toroidal magnetic field and endowed with Ohmic resistivity, the latter sufficient to suppress the MRI initially. We witnessed the exponential growth of magnetic energy and subsequent saturation of the dynamo in simulations with various values of the imposed cooling time and magnetic Reynolds number. Our main results can be summarized as follows.
\begin{enumerate}
\item During its kinematic phase, the GI dynamo grows a large-scale and predominantly toroidal magnetic field. It is a global process facilitated by a substantial amount of resistivity ($\Rm \lesssim 10$), though it may also operate in the ideal MHD limit using the effective resistivity of GI turbulence. 
\item The feedback loop leading to runaway magnetic amplification is the same as in the local simulations of \citet{riols19}, for which vertical (out-of-plane) motions are crucial. The vertical shear resulting from a global thermal stratification typically opposes the dynamo. 
\item The kinematic dynamo can be described via simple mean-field models upon azimuthal averaging. While its mechanism obeys the classic stretch-twist-fold terminology, its effect on the mean fields formally differs from traditional $\alpha$-$\Omega$ models.
\item Magnetic amplification stops near equipartition of the magnetic and turbulent energy densities. A saturated state of strong magnetization (i.e., plasma $\beta\sim 1$) is likely a robust result of the dynamo mechanism, and will be shared by both circumstellar and AGN settings. Here the magnetic stress becomes the primary source of turbulent energy, heating the disk faster than it can cool and bringing it away from GI, in agreement with \citet{deng20}. 
\end{enumerate}

Our focus through most of this paper was on the kinematic (linear) phase of the GI dynamo. Although our simulations reach nonlinear dynamo saturation, we believe that this final phase is largely determined by boundary conditions and limited integration times, and hence predictions regarding actual astrophysical objects might be premature at this point. The heat accumulating inside the computational domain may power outflows in more extended domains \citep{deng20}; the subsequent evacuation of magnetic flux may allow a new and hot steady state or overshoot and bring the disk back to the kinematic regime in a cyclic fashion \citep{riols19}. This calls for long-term simulations of the nonlinear dynamo phase in extended domains. 

Another route open for investigation concerns the microphysics. We have opted for the simplest modeling of the gas resistivity and radiative energy losses, fixing the magnetic Reynolds number $\Rm$ and cooling time $\tau$ to be constant, though in real disks both vary. On the one hand, a realistic treatment of radiative transfer can help clarify the role of GI-induced magnetic fields on disk fragmentation \citep{deng21}. On the other, circumstellar disks are notoriously nonideal electrical conductors, and the existence and properties of a GI dynamo should be further explored in realistic Hall and ambipolar MHD regimes \citep{riols21}.

\begin{acknowledgements}
  WB acknowledges funding by the Deutsche Forschungsgemeinschaft (DFG) through Grant KL 650/31-1, an instructive discussion with Yannick Ponty about numerical dynamo experiments, and a most uplifting visit to the DAMTP. 
  HNL acknowledges funding from STFC grant ST/T00049X/1.
  The authors acknowledge support by the High Performance and Cloud Computing Group at the Zentrum f\"ur Datenverarbeitung of the University of T\"ubingen, the state of Baden-W\"urttemberg through bwHPC and the DFG through grant no INST 37/935-1 FUGG.
  They also thank the anonymous reviewer, Antoine Riols, and François Rincon for their constructive feedback on the submitted version of the paper. 
  The preparation of this paper was overshadowed by mentor and friend Willy Kley's passing away; we carried his positive attitude throughout.
\end{acknowledgements}

\bibliographystyle{aa}
\bibliography{biblio}
  
\begin{appendix} 

  \section{Choice of EMF reconstruction scheme} \label{app:uct}

  Integrating the magnetic induction Eq. \eqref{eqn:dtB} over a bounded surface $\mathcal{S}$ and applying Stokes' theorem gives
  \begin{equation}
    \partial_t \int_{\mathcal{S}} \bm{B}\cdot \mathrm{d} \bm{S} = - \oint_{\partial\mathcal{S}} \bm{E} \cdot \mathrm{d} \bm{\ell},
  \end{equation}
  where $\bm{E}$ is the total EMF as seen in the chosen reference frame. The method of constrained transport implements this equation by taking the surface-integrated magnetic flux as variable to evolve in time \citep{evans88}. In a finite-volume framework, the different components of the discretized magnetic field therefore live on their respective interfaces and can be called ``face-centered''. On the other hand, the circulation of the EMF involves ``edge-centered'' variables.

  There are different ways to reconstruct the edge-centered EMF given the values of the face-centered and cell-centered MHD variables, with different accuracy and stability properties \citep{balsara99,londrillo04,gardiner05}. Throughout this paper we used the \verb|UCT_HLL| scheme of \citet{zanna03} implemented in \textsc{Pluto} 4.3. However, we also tried the less dissipative \verb|UCT0| and \verb|UCT_CONTACT| schemes of \citet{gardiner05} and observed spurious magnetic amplification in both cases. 

\begin{figure}
    \centering
    \includegraphics[width=\columnwidth]{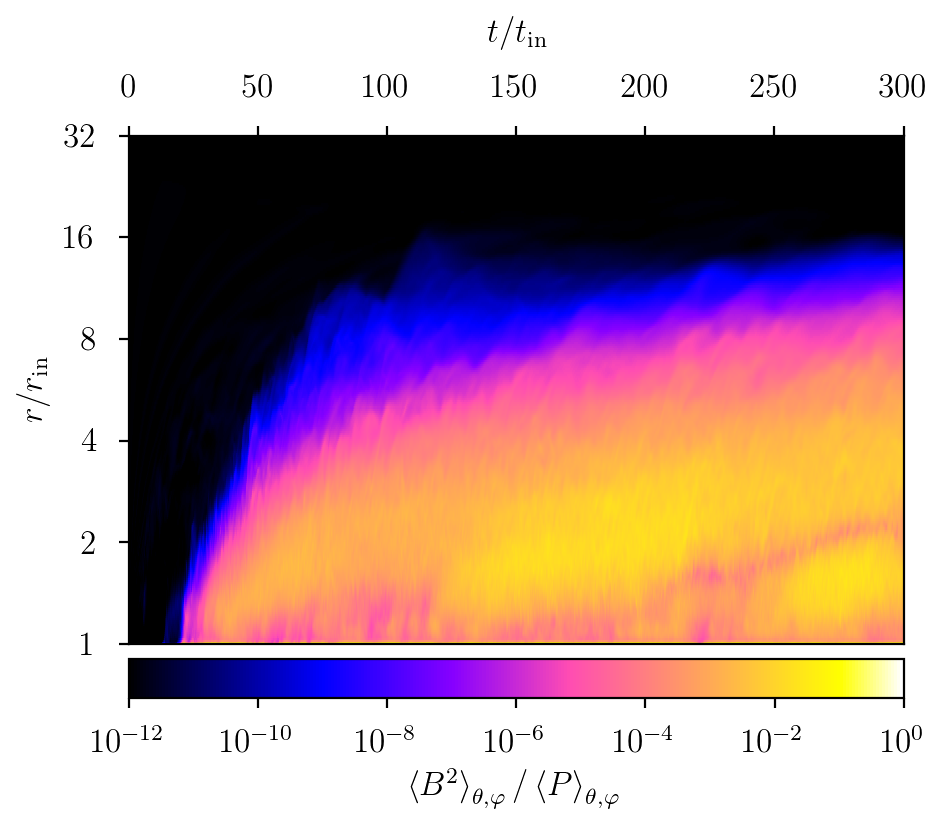}
    \caption{Evolution of the average magnetization ratio $\brac{B^2}/\brac{P}$ in radius (vertical axis) and time (horizontal axis) in a simulation equivalent to our reference case M3T10R10 (see Fig. \ref{fig:m3b10r10_map_b2}) but using the \texttt{UCT0} reconstruction scheme. This ratio increases by more than six orders of magnitude in less than ten inner orbits at time $t/t\lin \approx 25$.}
    \label{fig:m3b10r10uct0_map_b2}
  \end{figure}

Figure \ref{fig:m3b10r10uct0_map_b2} shows how the average disk magnetization evolves over time in a simulation equivalent to our reference case (see Fig. \ref{fig:m3b10r10_map_b2}) but using the \verb|UCT0| reconstruction scheme. After about $25 t\lin$, the ratio $B^2/P$ increases by more than six orders of magnitude in less than ten inner orbits. The corresponding growth rate is too fast to be associated with any fluid dynamo, leaving boundary effects as the only possibility. 

  We set the tangential components of the magnetic field to zero in the ghost cells, and we verified that there is no tangential magnetic flux injection appearing in the meridional profiles of $\bm{b}$ as defined by \eqref{eqn:redvar} and measured at time $100t\lin$. However, it is the $\nabla\cdot\bm{B}=0$ condition that controls the component normal to the boundaries. We found that the average proportion of toroidal field drops to $\brac{B_\varphi}_{\theta,\varphi} / \bar{B} \approx 0$ as soon as magnetic energy goes up. We deduce that magnetic energy is injected in $B_\theta$ fluctuations from the meridional boundaries, and confirm this after examining $B_\theta^2$ and $\partial_t B_\theta^2$ in the equivalent of Fig. \ref{fig:m3b10r10_EMF_DTBT_dtbt2}.

  If magnetic energy was injected from the domain boundaries in our simulations with \verb|UCT_HLL|, it did so with no significant consequence. Interestingly, the magnetization saturates below $B^2/P \lesssim 10^{-2}$ in Fig. \ref{fig:m3b10r10uct0_map_b2}, and the reduced magnetic field $\bm{b}$ displays the same meridional profiles as in Fig. \ref{fig:m3b10r10_EMF_B_extended}. These two points suggest that the GI dynamo may still be operating underneath the boundary-driven amplification, though at its slower pace.
  
  \section{Minimal mean-field instability} \label{app:mfd}

\begin{figure}
    \centering
    \includegraphics[width=\columnwidth]{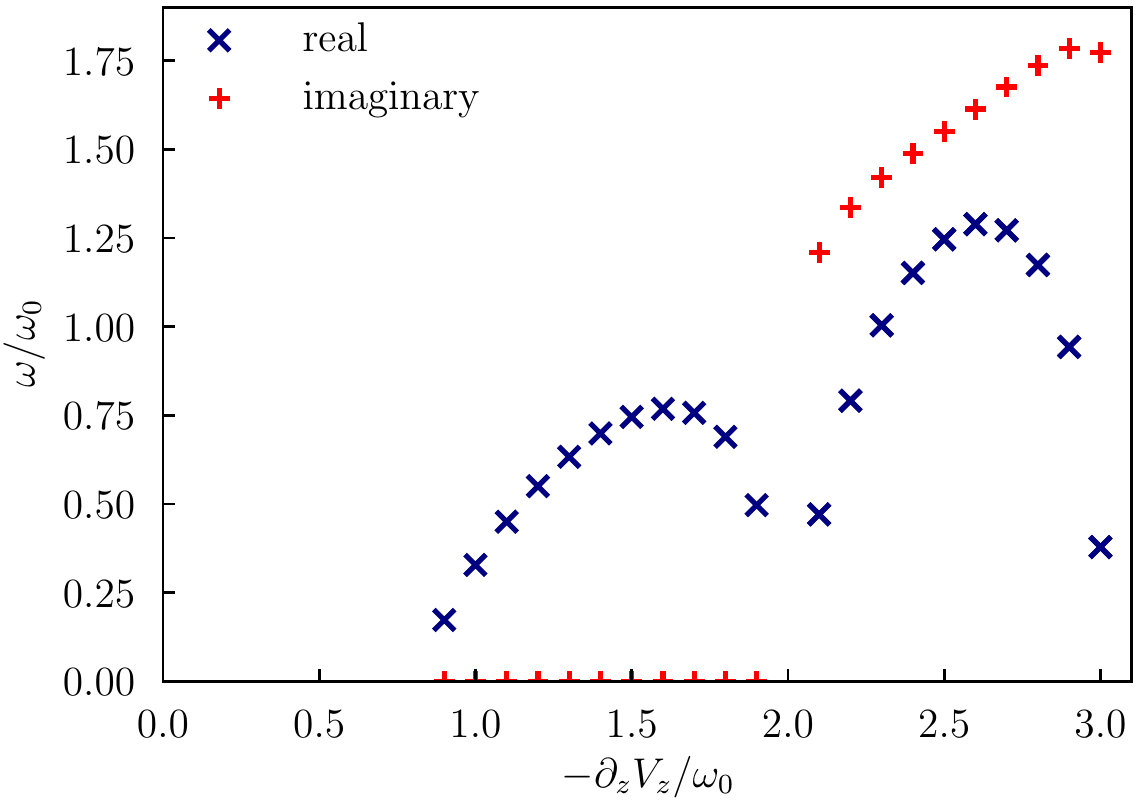}
    \caption{Eigenvalues of Eqs. \eqref{eqn:mfd3_dtbr} \& \eqref{eqn:mfd3_dtbp} for unstable modes with $\eta^{\prime}=1$ over $z\in \left[0,1\right]$, satisfying the conditions described in Sect. \ref{sec:mfd_eig}, and under different compression rates $-\partial_z V_z$ relative to the fundamental mode frequency $\omega_0$ obtained at $V_z=0$; blue (resp. red) crosses represent the real (resp. imaginary) part of the eigenvalue.}
    \label{fig:spidy_eigvals}
  \end{figure}

Equations \eqref{eqn:mfd2_dtbr} \& \eqref{eqn:mfd2_dtbp} can be further reduced to the minimal form:
\begin{align}
  \partial_t B_R \simeq &-\partial_z \left[ V_z B_R\right] - \eta^{\prime} \partial_z^2 B_\varphi \label{eqn:mfd3_dtbr},\\
  \partial_t B_\varphi \simeq &-\partial_z \left[ V_z B_\varphi\right] + \eta^{\prime} \partial_z^2 B_R, \label{eqn:mfd3_dtbp}
\end{align}
where, for simplicity, we took the same off-diagonal resistivities $\eta^{\prime}>0$ in both equations. We can then set $\eta^\prime=1$ and focus on $z\in\left[0,1\right]$ by rescaling space and time without loss of generality. We look for eigenmodes satisfying the conditions described in Sect. \ref{sec:mfd_eig} under a prescribed velocity $V_z\propto z$. In the case of $V_z=0$, the fundamental mode has a purely imaginary eigenvalue with frequency $\omega_0 \approx 5.59$. Unstable (real positive) eigenvalues appear when the amplitude of the compression rate $\vert\partial_z V_z\vert$ becomes comparable to the fundamental frequency $\omega_0$, as illustrated in Fig. \ref{fig:spidy_eigvals}. For larger values of the compression rate, unstable eigenvalues acquire a nonzero imaginary part and therefore correspond to growing oscillations --- similar to Fig. \ref{fig:m3b10r31_map_b2}.

\end{appendix}

\end{document}